\documentclass[11pt,onecolumn,a4paper]{article}

\usepackage[left=2.0cm,right=2.0cm,top=2.2cm,bottom=2.2cm]{geometry}
\usepackage[affil-it]{authblk}
\usepackage[utf8]{inputenc}
\usepackage{amsmath,amssymb,amsfonts,amsthm}
\usepackage{mathtools}
\usepackage{bbm}
\usepackage{microtype}
\usepackage{xcolor}
\usepackage{hyperref}
\usepackage[nameinlink]{cleveref}

\hypersetup{colorlinks=true, linkcolor=blue, citecolor=blue, urlcolor=cyan}
\newcommand{\GZ}{\mathrm{GZ}}
\newcommand{\RGZ}{\mathrm{RGZ}}

\newcommand{\gf}{\mathrm{gf}}
\newcommand{\R}{\mathbb{R}}
\renewcommand{\d}{\mathrm{d}}
\newcommand{\Int}{\int\!\d^{4}\!x\,}
\newcommand{\sdet}{\operatorname{sdet}}
\newcommand{\str}{\operatorname{str}}
\newcommand{\W}{\mathcal{W}}
\newcommand{\Msym}{\mathcal{S}}
\newcommand{\Masym}{\mathcal{A}}
\newcommand{\Bmat}{\mathcal{B}}
\newcommand{\Hor}{\mathsf{H}}

\newtheorem{theorem}{Theorem}[section]
\newtheorem{proposition}[theorem]{Proposition}
\newtheorem{lemma}[theorem]{Lemma}
\newtheorem{corollary}[theorem]{Corollary}

\begin{document}

\title{Local field shifts and the Gribov-Zwanziger horizon functional}
\author{Daniel G. Tedesco}
\affil{\small ESEHL/PPGENT, Curitiba, PR, Brazil\\
\small \texttt{daniel.te@uninter.com}}
\date{}
\maketitle

\begin{abstract}
We study a Maggiore-Schaden-like re-centering of the Gribov-Zwanziger action using a smooth compactly supported profile. At fixed background and finite cutoff, under the stated coercivity and contour hypotheses, this transformation preserves the original horizon functional with no residual dependence on the profile. The extension to gauge-field integration requires additional domination assumptions. Near a self-adjoint first Gribov horizon, the response depends on both the lowest Faddeev-Popov eigenvalue and its source weight. We extend this visibility analysis to non-self-adjoint compressions, separating source darkness from truncation-induced shielding, and recall continuum transverse $SU(2)$ examples. Transporting the composite insertion preserves the regulated Gaussian and resolvent structure. A separate free-space tree-level check reproduces the auxiliary-mass denominator of the RGZ propagator, without determining condensate dynamics.
\end{abstract}

\section{Introduction}
\label{sec:introduction}

Non-Abelian quantization still runs up against Gribov copies. Zero modes of the Faddeev-Popov operator signal the local breakdown of gauge fixing, while the topology of the orbit space rules out a globally unique representative~\cite{Gribov1978,Singer1978}. Variational methods allow us to identify the Faddeev-Popov operator with the Hessian of the gauge-fixing norm, distinguishing the first Gribov region from the fundamental modular region~\cite{SemenovFranke1986,DellAntonioZwanziger1991,vanBaal1992}. Restriction to the first Gribov region does not eliminate all Gribov copies, since every gauge orbit intersects this region under standard assumptions and positivity of the Hessian alone cannot eliminate all residual copies~\cite{LandimCopies2014,OrlandSemenoff2000}. We can explicitly construct normalizable zero modes for prescribed backgrounds using Henyey's inverse method~\cite{Henyey1979,GuimaraesSorella2011,CapriZeroModes2012,LandimZeroModes2014}.

To enforce the restriction to the first Gribov region, the Gribov-Zwanziger action introduces the auxiliary quartet $(\varphi,\bar\varphi,\omega,\bar\omega)$~\cite{Zwanziger1989,ZwanzigerLocal1989,Zwanziger1993,VandersickelZwanziger2012} that links the horizon and no-pole conditions and ensures the all-order renormalization of the horizon sector~\cite{CapriNoPole2013,DudalHorizon2010}. Besides, the theory naturally yields decoupling-type infrared propagators~\cite{Dudal2008,Dudal2009,Dudal2011} by "upgrading" to the refined formulation with dimension-two condensates, aligned with large-volume lattice simulations~\cite{CucchieriMendesGluon2008,Bogolubsky2009}. So, the BRST structure of this formulation has also been studied in terms of soft breaking, nonlocal $\gamma$-dependent symmetries, and nonperturbative formulations with a nilpotent differential~\cite{BaulieuSorella2009,Sorella2009,CapriGZ2010,CapriGZ2011,DudalSorella2012,Capri2013,Capri2015,Capri2016,Mintz2019}.

An alternative implementation of the Gribov restriction was proposed by Maggiore and Schaden~\cite{MaggioreSchaden1994}, in which the BRST transformations are recentered by field shifts proportional to $x_\mu$ and the horizon condition appears as a gap equation selecting a nontrivial vacuum, thereby giving the BRST breaking associated with the horizon a spontaneous interpretation. However, the coordinate dependence raises domain and boundary condition problems in the functional integral~\cite{VandersickelZwanziger2012,DudalSorella2012}.

We study a Maggiore-Schaden-like reparametrization of the local GZ action, which already contains the horizon coupling and vacuum term. Replacing $x_\mu$ with a smooth compactly supported profile keeps the displacement within the scalar Dirichlet domain and separates the change of variables from the question of BRST breaking. The gauge-fixing and auxiliary sectors have a BRST-exact continuum representation, whereas the full action in \eqref{eq:acao-GZ} is softly broken. In the refined theory, the operator $\bar\varphi\varphi-\bar\omega\omega$ must be carried along under the same field redefinition~\cite{Dudal2009,SchadenZwanziger2015a,SchadenZwanziger2015b}.

At fixed background and finite cutoff, the recentering leaves the horizon functional unchanged under the coercivity and contour hypotheses stated below. Extension to gauge-field integration and generating functionals is conditional on the additional trace, domination, and source-uniformity assumptions. Near a self-adjoint first Gribov horizon, a critical mode is weighted by its overlap with the Zwanziger source, as in spectral analyses of the ghost propagator~\cite{CucchieriMendesGhost2008,CucchieriMendes2013}. These are configurationwise statements and do not determine the statistical weight of dark configurations or contradict the ensemble horizon condition. Transporting $\Sigma=\bar\varphi\varphi-\bar\omega\omega$ preserves the regulated resolvent structure. The comparison at $J=-M^2$ with the tree-level RGZ propagator is a bare free-space consistency check, not a determination of the condensate or its effective potential~\cite{Dudal2008,Dudal2009,Dudal2011}.

We develop an admissible compact-profile recentering, extend it to coercive non-self-adjoint compressions, and derive the transported composite-source response. Building on the self-adjoint visibility criterion and rank classification of Ref.~\cite{TedescoDarkCrossings}, Proposition~2.2, the block analysis identifies the critical directions that survive an antisymmetric compression defect and characterizes their source visibility. Under the stated analytic hypotheses, this description yields bounded Laurent remainders and explicit source-rate conditions. The four-dimensional selection rule and smooth-shell thresholds established in Theorems~5.1--5.2 of that reference supply the continuum examples discussed below. Gaussian telescoping and inverse-moment identities follow algebraically from the construction, while the tree-level propagator and one-loop gap normalization provide consistency checks.

\section{Local action and regulated field spaces}
\label{sec:setting}

We use the gauge group $SU(N_c)$ and reserve $N$ for the Galerkin cutoff. In Euclidean signature the covariant derivative and Faddeev-Popov operator are
\begin{equation}
D_\mu^{ab}=\delta^{ab}\partial_\mu+g f^{acb}A_\mu^c,
\qquad
\mathcal M^{ab}[A]=-\partial_\mu D_\mu^{ab}[A].
\label{eq:covariant-fp}
\end{equation}
With these conventions, the local Gribov-Zwanziger action is~\cite{ZwanzigerLocal1989,Zwanziger1993}
\begin{align}
S_{\GZ}
&= \frac14\Int F_{\mu\nu}^aF_{\mu\nu}^a
 +\Int\!\left[ib^a\partial_\mu A_\mu^a
 +\bar c^a\partial_\mu D_\mu^{ab}c^b\right] \notag\\
&\quad
 +\Int\!\left[
 \bar\varphi_\mu^{ac}\partial_\nu D_\nu^{am}\varphi_\mu^{mc}
 -\bar\omega_\mu^{ac}\partial_\nu D_\nu^{am}\omega_\mu^{mc}
 -g(\partial_\nu\bar\omega_\mu^{ac})f^{a\ell m}(D_\nu c)^\ell\varphi_\mu^{mc}
 \right] \notag\\
&\quad
 -\gamma^2g\Int f^{abc}A_\mu^a
 \left(\varphi_\mu^{bc}+\bar\varphi_\mu^{bc}\right)
 -d(N_c^2-1)\gamma^4V.
\label{eq:acao-GZ}
\end{align}
The regulated identities are formulated at $d=4$. Dimensional continuation is used only for the one-loop gap equation, where the $d$ dependence of the vacuum term must be retained through subtraction. In the untruncated theory the localizers form the BRST doublets
\begin{align}
s\varphi_\mu^{ab}&=\omega_\mu^{ab},
& s\omega_\mu^{ab}&=0,
\notag\\
s\bar\omega_\mu^{ab}&=\bar\varphi_\mu^{ab},
& s\bar\varphi_\mu^{ab}&=0,
\label{eq:brst-localizers}
\end{align}
and satisfy $\Sigma=s(\bar\omega\varphi)$. Nonlinear BRST products need not close on a finite Galerkin space, so this continuum algebra does not define a finite-$N$ identity for the interacting regulated integral. Soft and linear breaking, as well as an extension with an exact nilpotent differential, provide distinct realizations of the horizon coupling~\cite{BaulieuSorella2009,CapriGZ2010,CapriGZ2011,Capri2014,DudalSorella2012}. The Dudal--Sorella extension is cited for context and is not used in the derivations below, which start from the softly broken local GZ action~\eqref{eq:acao-GZ}. Its additional fields and domain prescriptions lie outside the present construction. The regulated recentering establishes neither a finite-$N$ Slavnov--Taylor identity nor spontaneous BRST breaking.

Inside the first Gribov region, $\mathcal M$ is positive and the bosonic localizer integral must be defined on a convergent contour. Since we use $K=-\mathcal M$, we write
\begin{equation}
\varphi=\frac{P+iQ}{\sqrt2},
\qquad
\bar\varphi=-\frac{P-iQ}{\sqrt2},
\qquad P,Q\ \text{real},
\end{equation}
so that, for symmetric $\mathcal M$,
$$
\bar\varphi K\varphi
=
\frac12\langle P,\mathcal MP\rangle
+
\frac12\langle Q,\mathcal MQ\rangle .
$$
The minus sign in $\bar\varphi$ compensates for the convention $K=-\mathcal M$; with $\bar\varphi=\varphi^*$ the same kernel would give a divergent Gaussian. We absorb this sign by defining
\begin{equation}
\widehat\varphi=\varphi,
\qquad
\widehat{\bar\varphi}=-\bar\varphi,
\qquad
\widehat m=m,
\qquad
\widehat{\bar m}=-\bar m ,
\label{eq:dictionary-map}
\end{equation}
which restores $\widehat{\bar\varphi}=\widehat\varphi^{,*}$ without changing $(P,Q)$ or the measure $\d P,\d Q$. The localizer integral then has the usual Gaussian form,
\begin{equation}
\int\d P,\d Q,
e^{-\widehat{\bar\varphi}\mathcal M\widehat\varphi
-\widehat{\bar\varphi}\widehat{\bar m}
-\widehat m^{\mathsf T}\widehat\varphi}
\propto
(\det\mathcal M)^{-1}
\exp\!\left[
+\widehat m^{\mathsf T}\mathcal M^{-1}\widehat{\bar m}
\right],
\label{eq:dictionary-gaussian}
\end{equation}
and therefore
\begin{equation}
\Delta S_{\rm hor}
=-m^{\mathsf T}K^{-1}\bar m
=m^{\mathsf T}\mathcal M^{-1}\bar m
=\gamma^4\Hor(A).
\label{eq:dictionary-horizon}
\end{equation}
For $m=\bar m$ and $\mathcal M>0$, this is the standard nonnegative horizon pairing. The Grassmann determinant cancels $(\det\mathcal M)^{-1}$, leaving the expected factor $e^{-\gamma^4\Hor}$ without changing the real localizer measure~\cite{VandersickelZwanziger2012,Dudal2008}.

We regularize the theory on the ball
\begin{equation}
B_R=\{x\in\R^4:|x|<R\},
\qquad R>2\Lambda,
\label{eq:ball}
\end{equation}
using complete eigenspaces of the Dirichlet Laplacian below the cutoff $\Lambda_{\rm UV}$,
\begin{equation}
\mathcal H_N
=\bigoplus_{\lambda_r\leq\Lambda_{\rm UV}^2}
\operatorname{span}\{\psi_r\},
\qquad
P_N:L^2(B_R)\longrightarrow\mathcal H_N,
\qquad
d_N=\dim\mathcal H_N .
\label{eq:spectral-space}
\end{equation}
To preserve $O(4)$ symmetry at finite cutoff, we retain complete degenerate multiplets of the Laplacian. The Faddeev-Popov ghosts, the $b$ field, and the localizing fields are truncated to $\mathcal H_N$, while $A_\mu$ is taken in a finite-dimensional vector space $\mathcal V_N$, chosen so that the projected divergence, the regulated Faddeev-Popov operator, and the required integrations by parts are well defined. When needed, $\mathcal H_N$ and $\mathcal V_N$ are chosen to be divergence-compatible, so that the projected Landau condition reproduces transversality. Redundant constraints are removed before taking the Landau limit.

Color factors are implicit in the scalar-space notation, and $P_N$ acts componentwise on adjoint fields. Define
\begin{equation}
F_N[A]=P_N\partial_\mu A_\mu,
\qquad
\mathcal M_N[A]=P_N\bigl(-\partial_\mu D_\mu[A]\bigr)P_N,
\qquad
K_N=-\mathcal M_N .
\label{eq:fp-regulated}
\end{equation}
\paragraph{A compatible Dirichlet realization.}
Let $\{\psi_r\}$ be the retained orthonormal scalar eigenfunctions, with $-\partial^2\psi_r=\lambda_r\psi_r$ and $\lambda_r>0$. Choose finitely many radial $f_k\in C_c^\infty(B_R)$ and let $\mathcal V_N^T$ be the span, componentwise in color, of $e_\mu\partial_\nu f_k-e_\nu\partial_\mu f_k$ for $\mu<\nu$, after removal of linearly dependent generators. These vectors are smooth, divergence-free, and supported away from the boundary. Taking all pairs $\mu<\nu$ preserves $O(4)$ covariance. Set
\begin{equation*}
\mathcal V_N=\mathcal V_N^T\oplus\nabla\mathcal H_N,
\qquad
A=A_T+\sum_r q_r\nabla\psi_r,
\qquad
F_N[A]=-\sum_r\lambda_rq_r\psi_r .
\end{equation*}
The scalar fields obey homogeneous Dirichlet conditions. The vector traces satisfy $A_{\rm tan}|_{\partial B_R}=0$ and $(\partial\!\cdot\!A)|_{\partial B_R}=0$, while the normal component of $A$ is not prescribed. These matched traces allow the gradient sector, which would generally be excluded by homogeneous Dirichlet conditions on every vector component. The sum is direct because a harmonic Dirichlet scalar is zero. The displayed diagonal divergence block is invertible, so $F_N$ is onto $\mathcal H_N$ and has full row rank, with no redundant scalar constraint. Its kernel is precisely $\mathcal V_N^T$, and $\operatorname{div}\mathcal V_N\subseteq\mathcal H_N$ gives exact transversality on the Landau fiber. The integrations by parts below have no boundary contribution because their scalar test functions vanish on $\partial B_R$.

Projection of every shifted field coefficient keeps the affine map in these spaces. At fixed $A$ with $\Msym_N[A]\geq\delta I$, after the regulated multiplier integration, the finite Grassmann expansion leaves polynomial factors multiplying the localizer Gaussian. Bounded translations of $P,Q$ add only linear terms to its real exponent and finite constants, giving a bound $C_A(1+\|P\|+\|Q\|)^k\exp[-\delta(\|P\|^2+\|Q\|^2)/4]$ on the translation strips. This checks the reduced fixed-background contour requirement. The constant is not asserted to be integrable over all gauge coefficients, so the extension to gauge-field integration retains the hypotheses following \cref{thm:fiberwise}. Divergence compatibility is used for the self-adjoint spectral interpretation, while the reduced Gaussian argument only requires coercivity.

For Dirichlet test functions $v,w\in\mathcal H_N$, integration by parts and relabeling of the adjoint indices give
\begin{equation}
\langle v,\mathcal M_Nw\rangle
-\langle\mathcal M_Nv,w\rangle
=-g\Int f^{acb}(\partial_\mu A_\mu^c)\,v^aw^b .
\label{eq:antisymmetric-part}
\end{equation}
The condition $F_N[A]=0$ makes $\partial_\mu A_\mu$ orthogonal to $\mathcal H_N$, but the products $v^aw^b$ in this pairing need not lie in $\mathcal H_N$. Consequently, projected transversality can leave an adjoint defect. Writing
\begin{equation}
\mathcal M_N=\Msym_N+\Masym_N,
\qquad
\Msym_N=\tfrac12\bigl(\mathcal M_N+\mathcal M_N^{\mathsf T}\bigr),
\qquad
\Masym_N=\tfrac12\bigl(\mathcal M_N-\mathcal M_N^{\mathsf T}\bigr),
\label{eq:symmetric-split}
\end{equation}
we impose coercivity on the symmetric part through
\begin{equation}
\widetilde\Omega_{N,\delta}
=\left\{A:\Msym_N[A]\geq\delta\mathbbm 1\right\},
\qquad
\Omega_{N,\delta}
=\left\{A\in\widetilde\Omega_{N,\delta}:F_N[A]=0\right\},
\label{eq:omega-delta}
\end{equation}
for a fixed $\delta>0$. These domains do not impose the additional matrix constraints $\Masym_N=0$.

On the localizer cycle, the quadratic form reads
\begin{equation}
\langle\bar\varphi,K_N\varphi\rangle
=\tfrac12\langle P,\Msym_NP\rangle
+\tfrac12\langle Q,\Msym_NQ\rangle
+i\langle P,\Masym_NQ\rangle .
\label{eq:PQ-form}
\end{equation}
The antisymmetric part contributes a mixed phase, while the real part is controlled by $\Msym_N$. In particular,
\begin{equation}
\left|e^{-\langle\bar\varphi,K_N\varphi\rangle}\right|
\leq
\exp\!\left[-\frac{\delta}{2}\bigl(\|P\|^2+\|Q\|^2\bigr)\right]
\qquad\text{for }A\in\Omega_{N,\delta}.
\label{eq:uniform-localizer-bound}
\end{equation}
The phase changes the value of the Gaussian despite leaving this modulus bound unchanged. Integration over $P$ is nevertheless controlled by the symmetric part alone.

\begin{lemma}[Schur reduction of the localizer Gaussian]
\label{lem:schur}
Let $\Msym$ be real symmetric with $\Msym\geq\delta\mathbbm1$ and let $\Masym$ be real antisymmetric, and set
\begin{equation}
\Bmat=\Msym-\Masym\Msym^{-1}\Masym
=\Msym+\Masym^{\mathsf T}\Msym^{-1}\Masym .
\label{eq:Schur-B}
\end{equation}
Then $\Bmat$ is positive definite, and for every real source $m$,
\begin{align}
&\int_{\R^{2n}}\d P\,\d Q\,
\exp\!\left[
-\tfrac12P^{\mathsf T}\Msym P
-\tfrac12Q^{\mathsf T}\Msym Q
-iP^{\mathsf T}\Masym Q
-i\sqrt2\,Q^{\mathsf T}m
\right]
\notag\\
&\hspace{4.2cm}
=C(\Msym,\Masym)\,\exp\!\left[-m^{\mathsf T}\Bmat^{-1}m\right],
\label{eq:exact-PQ-gaussian}
\end{align}
with $C$ independent of $m$. The scalar form of the full inverse agrees with that of $\Bmat^{-1}$,
\begin{equation}
\tfrac12\left[(\Msym+\Masym)^{-1}+(\Msym-\Masym)^{-1}\right]=\Bmat^{-1},
\qquad
m^{\mathsf T}(\Msym+\Masym)^{-1}m=m^{\mathsf T}\Bmat^{-1}m>0
\label{eq:symmetric-inverse-identity}
\end{equation}
for $m\neq0$. Symmetry of the compressed operator is not used.
\end{lemma}

\begin{proof}
Positivity is the identity
\begin{equation}
v^{\mathsf T}\Bmat v
=v^{\mathsf T}\Msym v
+(\Masym v)^{\mathsf T}\Msym^{-1}(\Masym v),
\label{eq:B-positive}
\end{equation}
both terms being nonnegative and the first bounded below by $\delta\|v\|^2$. Completing the square in $P$ and then in $Q$ gives~\eqref{eq:exact-PQ-gaussian}. For the last statement, write $X=\Msym+\Masym$ and $Y=X^{\mathsf T}$. The identities $X\Msym^{-1}Y=Y\Msym^{-1}X=\Bmat$ give $\Bmat^{-1}=X^{-1}\Msym Y^{-1}=Y^{-1}\Msym X^{-1}$, and $X+Y=2\Msym$ then yields $X^{-1}+Y^{-1}=2\Bmat^{-1}$. Contracting with a real $m$ on both sides makes the two scalar forms coincide.
\end{proof}

The horizon pairing is therefore positive on the whole coercive domain, whether or not $\mathcal M_N$ is normal there.

For the recentering, choose a radial $\chi_\Lambda\in C_0^\infty(\R^4)$ with $0\leq\chi_\Lambda\leq1$, equal to one on $|x|\leq R_1(\Lambda)$ and supported in $B_\Lambda$. Set
\begin{equation}
X_\mu(x)=x_\mu\chi_\Lambda(x),
\qquad
u_\mu^{ab}(x)=\gamma^2\delta^{ab}X_\mu(x).
\label{eq:shift-u}
\end{equation}
The affine transformation is
\begin{align}
\varphi_\mu^{ab}
 &=\varphi_\mu^{\prime ab}+u_\mu^{ab},
&
\bar\varphi_\mu^{ab}
 &=\bar\varphi_\mu^{\prime ab}+u_\mu^{ab},
\label{eq:shift-bosons}\\
\bar c^a
 &=\bar c^{\prime a}+g\gamma^2f^{abc}\bar\omega_\mu^{bc}X_\mu,
&
b^a
 &=b^{\prime a}-ig\gamma^2f^{abc}\bar\varphi_\mu^{\prime bc}X_\mu,
\label{eq:shift-gf}
\end{align}
with all other fields unchanged. The $b$ shift is written in terms of $\bar\varphi'$ to display its triangular dependence. Replacing $\bar\varphi'$ by $\bar\varphi$ gives the same expression because $f^{abc}\delta^{bc}=0$. The factor $-i$ follows from $s\bar c=ib$, and substitution of $s\bar\omega=\bar\varphi$ and $sX=0$ gives $s\bar c'=ib'$ and $sb'=0$. In the hatted convention the bosonic shifts have opposite signs,
\begin{equation}
\widehat\varphi=\widehat{\varphi'}+u ,
\qquad
\widehat{\bar\varphi}=\widehat{\bar\varphi'}-u ,
\label{eq:dictionary-recentering}
\end{equation}
whereas the unchanged $\bar\omega$ satisfies
\begin{equation}
s\bar\omega_\mu^{ab}
=\bar\varphi_\mu^{\prime ab}+\gamma^2\delta^{ab}X_\mu(x)
\;\xrightarrow{\ \Lambda\to\infty\ }\;
\bar\varphi_\mu^{\prime ab}+\gamma^2\delta^{ab}x_\mu .
\label{eq:sbaromega-mod}
\end{equation}
At fixed $x$, $\partial_\nu X_\mu=\delta_{\nu\mu}\chi_\Lambda+x_\mu\partial_\nu\chi_\Lambda$ tends to $\delta_{\nu\mu}$ as the plateau expands. The coordinate-dependent Maggiore-Schaden algebra is recovered pointwise, and nilpotency on $\bar\omega$ is preserved because $s\bar\varphi'=0$ and the profile is BRST inert.

Ordering the fields so that the variables entering the shifts precede their images gives
\begin{equation}
\frac{\partial\Phi}{\partial\Phi'}=\mathbbm 1+N,
\qquad
N_{\bar c\,\bar\omega}=g\gamma^2f^{abc}X_\mu,
\qquad
N_{b\,\bar\varphi'}=-ig\gamma^2f^{abc}X_\mu.
\label{eq:jacobian-blocks}
\end{equation}
The off-diagonal blocks cannot compose, so $N^2=0$, and the absence of field-diagonal blocks gives $\str N=0$. Hence
\begin{equation}
\sdet\!\left(\frac{\partial\Phi}{\partial\Phi'}\right)
=\exp\!\left[\str\ln(\mathbbm 1+N)\right]
=\exp\str N=1.
\label{eq:sdet}
\end{equation}
Closure on the Galerkin spaces also requires projection of the profile and of every product in the transformation. With $X_{N,\mu}=P_NX_\mu$ and $u_{N,\mu}^{ab}=\gamma^2\delta^{ab}X_{N,\mu}$, the finite-dimensional map is
\begin{align}
\varphi_N&=\varphi_N'+u_N,
&
\bar\varphi_N&=\bar\varphi_N'+u_N,
\notag\\
\bar c_N&=\bar c_N'+P_N\!\left(g\gamma^2f\bar\omega_NX_N\right),
&
b_N&=b_N'-iP_N\!\left(g\gamma^2f\bar\varphi_N'X_N\right).
\label{eq:projected-map}
\end{align}
It retains the triangular Jacobian, but its admissibility in the integral additionally depends on the displaced bosonic contours.

\section{Admissible recentering and the horizon functional}
\label{sec:contours}

The multiplier integral is defined at nonzero gauge parameter,
\begin{equation}
S_{\gf,N,\alpha}
=i\langle b_N,F_N[A]\rangle
+\frac{\alpha}{2}\langle b_N,b_N\rangle,
\qquad \alpha>0 ,
\label{eq:b-regulator}
\end{equation}
so integration over its real coefficient space gives
\begin{equation}
\int_{\R^{n_b}}\frac{\d^{n_b}b}{(2\pi)^{n_b}}
\exp\!\left[-\frac{\alpha}{2}b^{\mathsf T}b-ib^{\mathsf T}F\right]
=\delta_\alpha(F),
\qquad
\delta_\alpha(F)=\frac{e^{-F^{\mathsf T}F/(2\alpha)}}{(2\pi\alpha)^{n_b/2}} .
\label{eq:smeared-delta}
\end{equation}
The Landau constraint is obtained from the distributional limit $\delta_\alpha\to\delta$ after this integration. At fixed spectator variables, a finite displacement $c\in\mathbb C^{n_b}$, including one that depends polynomially on those variables, satisfies
\begin{equation}
\int_{\R^{n_b}+ic}\d^{n_b}b\;
 e^{-\alpha b^{\mathsf T}b/2-ib^{\mathsf T}F}
=
\int_{\R^{n_b}}\d^{n_b}b\;
 e^{-\alpha b^{\mathsf T}b/2-ib^{\mathsf T}F}.
\label{eq:b-contour-shift}
\end{equation}
The integrand is entire. Translation invariance removes the real part of the displacement, and Gaussian decay on the strip swept by its imaginary part makes the vertical sides of the Cauchy rectangles vanish. This argument applies componentwise and leaves an integral independent of $c$.

The localizer deformation is controlled by the coercive symmetric kernel. For a polynomial $\mathcal P$ and finite spectator-dependent coefficients $\eta,w$ independent of $z$, consider
\begin{equation}
f(z)=\mathcal P(z)\exp\!\left[-\tfrac12z^{\mathsf T}\Msym z-i\eta^{\mathsf T}\Masym z-iw^{\mathsf T}z\right],
\qquad z\in\mathbb C^n,
\label{eq:strip-integrand}
\end{equation}
with $\Msym\geq\delta\mathbbm1$ and $\Masym^{\mathsf T}=-\Masym$. Translation of one component through $z=x+iv_je_j$, with $x$ real, gives
\begin{equation}
z^{\mathsf T}\Msym z
=x^{\mathsf T}\Msym x
+2iv_j(\Msym x)_j
-v_j^2\Msym_{jj},
\label{eq:strip-form}
\end{equation}
and therefore
\begin{equation}
\bigl|e^{-\frac12z^{\mathsf T}\Msym z}\bigr|
\leq e^{\frac12v_j^2\Msym_{jj}}e^{-\frac{\delta}{2}|x|^2}.
\label{eq:strip-bound}
\end{equation}
The linear terms contribute at most $e^{C_0+C_1|x|}$ on a finite strip. Combining $C_1|x|\leq\delta|x|^2/4+C_1^2/\delta$ with the polynomial factor gives an integrable bound $C(1+|x|)^k e^{-\delta|x|^2/4}$. Successive componentwise deformations thus establish
\begin{equation}
\int_{\R^n}\d^nQ\,f(Q)=\int_{\R^n}\d^nQ\,f(Q+iv),
\label{eq:Q-contour-shift}
\end{equation}
for every finite real $v$. The same estimate permits the finite complex translations of both $P$ and $Q$ needed to complete a Gaussian with independent left and right sources.

For the projected $b$ displacement, $c=P_N(g\gamma^2f\bar\varphi_N'X_N)$, the contour identity holds at each $\alpha>0$. Both parametrizations consequently give the same $\delta_\alpha(F_N[A])$, whose limit imposes $F_N[A]=0$. On this fiber the equal-sign localizer shift leaves $P$ fixed and translates $Q$ according to
\begin{equation}
P'=P,
\qquad
Q'=Q+i\sqrt2\,u_N .
\label{eq:PQ-translation}
\end{equation}
The mixed term then contributes $\exp[+\sqrt2\langle P,\Masym_Nu_N\rangle]$. Its exponent is linear in $P$ and is absorbed by the Gaussian bound at fixed $\delta>0$. Integration over spectator variables requires an integrable majorant for the resulting contour homotopy.

\begin{theorem}[Admissibility on the transverse fiber]
\label{thm:fiberwise}
Fix $(R,N,\Lambda,\delta)$ with $\delta>0$ and $A\in\Omega_{N,\delta}$. Suppose that $F_N$ has full row rank, the projected affine map closes on the regulated coefficient spaces, and its contour homotopies have a common integrable majorant on the finite strips they sweep. For every polynomial insertion independent of $b_N$, with fixed sources coupled to untranslated fields, the reduced localizer-Grassmann integrals satisfy
\begin{equation}
\mathfrak I^{\rm transl}_{N,R,\Lambda,\delta}[\mathcal O;A]
=
\mathfrak I^{\rm unshifted}_{N,R,\Lambda,\delta}[\mathcal O;A],
\label{eq:finite-equivalence}
\end{equation}
where the translated integral contains the pullback of both the action and the insertion.
\end{theorem}

\begin{proof}
The projected map has unit superdeterminant. At fixed localizer and ghost coefficients, the multiplier contour returns to the real cycle before the Landau limit, giving the same smeared delta in both parametrizations. Full row rank allows the common coarea factor to be removed, leaving densities on $F_N[A]=0$. The finite localizer translation is admissible by the strip estimate and the assumed majorant, while the Grassmann translation has unit Berezinian. Projection keeps every coefficient in the retained space, so these changes of variables yield the equality of reduced integrals.
\end{proof}

The fiber identity extends to gauge-field integration when the pre-Landau measure on $\widetilde\Omega_{N,\delta}$ admits disintegration along $F_N$, its integrated densities have a common continuous trace at $F_N=0$, and a common integrable bound permits the approximate-delta limit, contour deformation, and background integration. Under these conditions the approximate delta converges to the trace and the fiber equality can be integrated over $A$. Insertions containing $b_N$ produce derivatives of the Landau delta and require a separate distributional treatment.

On a common source neighborhood, source-uniform domination and differentiability give equality of generating functionals for sources coupled to untranslated fields. If $Z[J]\neq0$ and the same branch of $\log Z[J]$ is used, their connected correlators also coincide. Uniformity under sufficiently small admissible profile variations $X_N+t\eta_N$ then gives
\begin{equation}
\left.\frac{\d}{\d t}W_{N,R,\Lambda,\delta}[X_N+t\eta_N]\right|_{t=0}=0 ,
\label{eq:profile-redundancy}
\end{equation}
by differentiation under the integral. The contour estimate remains valid at $\lambda_{\min}(\Msym_N)=\delta>0$. Uniform control of the localizer strip estimates additionally requires fixed coefficient-space dimension, uniformly bounded kernel and source data, and bounded contour displacements. With $(R,N,\Lambda)$ and $\gamma$ fixed, these bounds hold on compact parameter sets with a common positive coercive lower bound. The estimates do not establish uniformity under regulator removal or unrestricted background integration. The multiplier contour identity is used at each fixed $\alpha>0$, and the Landau limit retains the separate domination assumptions stated above.

At fixed gap, the localizer Gaussian evaluates the horizon contribution without an expansion in $A$. Suppressing regulator labels, with all operators and source columns understood on the retained coefficient spaces, the right and left sources, coupled respectively to $\bar\varphi'$ and $\varphi'$, are
\begin{align}
\bar m_\mu^{ac}&=+\gamma^2g f^{a\ell c}A_\mu^\ell
,
\label{eq:fonte-mbar}\\
m_\mu^{ma}&=-\gamma^2g f^{a\ell m}A_\mu^\ell
.
\label{eq:fonte-m}
\end{align}
The arrays coincide because $\bar m_\mu^{ma}=\gamma^2gf^{m\ell a}A_\mu^\ell=-\gamma^2gf^{a\ell m}A_\mu^\ell=m_\mu^{ma}$. Color antisymmetry also gives
\begin{equation}
u^{\mathsf T}\bar m
=m^{\mathsf T}u
=\gamma^4g\Int f^{a\ell a}X_\mu A_\mu^\ell=0 ,
\label{eq:shift-source-orthogonality}
\end{equation}
and removes the translated part of the horizon coupling. The bosonic action consequently becomes
\begin{equation}
S_B'
=\bar\varphi'K\varphi'
+\bar\varphi'\,a
+\ell^{\mathsf T}\varphi'
+u^{\mathsf T}Ku ,
\qquad
a=Ku+\bar m,
\qquad
\ell^{\mathsf T}=u^{\mathsf T}K+m^{\mathsf T}.
\label{eq:combined-sources}
\end{equation}
The profile terms and horizon sources belong to the same Gaussian pairing.

\begin{theorem}[Horizon telescoping]
\label{thm:telescoping}
For $A\in\Omega_{N,\delta}$ at fixed $(R,N,\Lambda,\delta)$, on an admissible contour as in \cref{thm:fiberwise}, integration over the translated localizers gives
\begin{equation}
\Delta S_{\rm hor}
=-m^{\mathsf T}K^{-1}\bar m
=m^{\mathsf T}\mathcal M^{-1}\bar m
=\gamma^4\Hor(A),
\label{eq:horizon-telescoping}
\end{equation}
with the original coefficient of the regulated horizon functional and no dependence on the profile. Symmetry of $K$ is not required.
\end{theorem}

\begin{proof}
Completion of the square on the admissible contour gives
\begin{equation}
\Delta S_{\rm hor}=-\ell^{\mathsf T}K^{-1}a+u^{\mathsf T}Ku .
\label{eq:raw-telescoping}
\end{equation}
Using both inverse identities for $K$,
\begin{align}
\ell^{\mathsf T}K^{-1}a
&=\bigl(u^{\mathsf T}K+m^{\mathsf T}\bigr)K^{-1}\bigl(Ku+\bar m\bigr)
\notag\\
&=u^{\mathsf T}Ku+u^{\mathsf T}\bar m+m^{\mathsf T}u+m^{\mathsf T}K^{-1}\bar m
=u^{\mathsf T}Ku+m^{\mathsf T}K^{-1}\bar m .
\label{eq:expanded-telescoping}
\end{align}
The mixed contractions vanish by color antisymmetry, and the two quadratic profile terms cancel. Since $K^{-1}=-\mathcal M^{-1}$, the remaining term is \eqref{eq:horizon-telescoping}. For the coincident real source arrays, positivity follows from \eqref{eq:symmetric-inverse-identity}.
\end{proof}

For smooth uncompressed fields on a fixed ball, the same cancellation has a perturbative channel representation. With the compact profile $X_\mu=x_\mu\chi_\Lambda$ and $K=K_0+K_1$, where $K_0=\partial^2$ and $K_1=gf\,\partial_\nu(A_\nu\,\cdot)$, direct application to the shift gives
\begin{equation}
(Ku)_\mu^{ac}
=\gamma^2\delta^{ac}\partial^2X_\mu
+\gamma^2g f^{a\ell c}
\left[(\partial\!\cdot\!A^\ell)X_\mu
+A_\mu^\ell\chi_\Lambda
+(A^\ell\!\cdot\!\partial\chi_\Lambda)x_\mu\right].
\label{eq:Ku-exact}
\end{equation}
The Laplacian term is supported where the window varies. The term proportional to $(\partial\!\cdot\!A)X$ cancels the coordinate-dependent contribution from the multiplier shift after relabeling $f^{bac}=-f^{abc}$. In the ghost sector, integration by parts cancels the terms proportional to $X\partial_\nu(D_\nu c)$. On the transverse plateau,
\begin{equation}
(Ku)_\mu^{ac}\big|_{\rm plateau,\ \partial\cdot A=0}=\gamma^2gf^{a\ell c}A_\mu^\ell=\bar m_\mu^{ac},
\qquad
K_1^{\mathsf T}u=m ,
\label{eq:plateau-collapse}
\end{equation}
where the transpose identity also uses integration by parts. If $\operatorname{supp}A$ lies strictly inside the plateau, both identities hold throughout the interaction support and can be used inside the nonlocal pairing.

Expanding about $K_0$, under the convergence condition $\rho(K_0^{-1}K_1)<1$, and using $K_0^{-1}(K_0u)=u$, the terms of order $A^2$ are
\begin{align}
\text{(i)}\quad
\bigl(m+K_1^{\mathsf T}u\bigr)^{\mathsf T}K_0^{-1}\bigl(\bar m+K_1u\bigr)
&\longrightarrow+4\,m^{\mathsf T}K_0^{-1}\bar m,
\label{eq:channel-i}\\
\text{(ii)}\quad
-u^{\mathsf T}K_1K_0^{-1}\bigl(\bar m+K_1u\bigr)
-\bigl(m+K_1^{\mathsf T}u\bigr)^{\mathsf T}K_0^{-1}K_1u
&\longrightarrow-4\,m^{\mathsf T}K_0^{-1}\bar m,
\label{eq:channel-ii}\\
\text{(iii)}\quad
+u^{\mathsf T}K_1K_0^{-1}K_1u
&\longrightarrow+1\cdot m^{\mathsf T}K_0^{-1}\bar m .
\label{eq:channel-iii}
\end{align}
Their coefficients sum to $4-4+1=1$. The first channel alone would therefore overcount the horizon self-energy by a factor of four, because the other two retain the contractions involving $K_0u$. Without the support assumption, the channel expressions still cancel algebraically, but their separate coefficients cannot be replaced by $4,-4,1$. Finite projection can also invalidate the product rules used in this smooth-field decomposition.

Compact support supplies the inverse-domain condition $K_{0,D}^{-1}K_0u=u$ on the finite Dirichlet ball, where $K_{0,D}^{-1}$ is the Dirichlet Green operator. This identity justifies the preceding cancellations but does not replace that Green operator by a Fourier multiplier in a general source pairing. For a separate free-space tree-level consistency check, take the translation-invariant Green operator on $\R^4$, represented by $-1/p^2$. Then $\tilde X_\mu(p)=i\partial_{p_\mu}\tilde\chi_\Lambda(p)$ and $\widetilde{K_0u}(p)=-p^2\tilde u(p)$ give the same cancellation for $p\neq0$. Contracting $f^{a\ell m}f^{a\ell'm}=N_c\delta^{\ell\ell'}$ yields the self-energy $2N_cg^2\gamma^4/p^2$ and transverse inverse propagator $p^2+\lambda^4/p^2$, with $\lambda^4=2N_cg^2\gamma^4$. This momentum-space check assumes the free-space prescription and does not assert convergence of the regulated Dirichlet integral.

The unregulated coordinate function $x_\mu$ belongs neither to the homogeneous Dirichlet domain on $B_R$ nor to $L^2(\R^4)$.

The Gaussian contraction also cancels the translated quadratic constant,
\begin{equation}
u^{\mathsf T}Ku=-\gamma^4(N_c^2-1)\sum_{\mu,\nu}\int\!\d^4x\,(\partial_\nu X_\mu)^2,
\label{eq:profile-quadratic-constant}
\end{equation}
so the vacuum term in the original action is unchanged. Compact support independently implies
\begin{equation}
\Int\partial_\nu X_\nu
=4\Int\chi_\Lambda-\Int(\partial_\nu x_\nu)\chi_\Lambda
=4\Int\chi_\Lambda-4\Int\chi_\Lambda=0 ,
\label{eq:profile-divergence-zero}
\end{equation}
whereas substitution of $X_\mu=x_\mu$ before integration gives $\partial_\mu x_\mu=4$. Extending each compact profile by zero on $\R^4$, with plateaux that exhaust the space, therefore yields
\begin{equation}
\lim_{\Lambda\to\infty}\Int\partial_\nu X_\nu=0
\;\neq\;
\Int\lim_{\Lambda\to\infty}\partial_\nu X_\nu=+\infty .
\label{eq:vacuum-noncommutativity}
\end{equation}
On a fixed ball, $X=x$ would instead give $4V$ by changing the boundary data, not by taking the regulated limit $R>2\Lambda$. The compact construction leaves the single vacuum term $-d(N_c^2-1)\gamma^4V$, whose one-loop normalization gives the gap-equation constant $5/3$~\cite{Gracey2006}.

Transport of the one-point function gives $\langle\bar\varphi'\rangle=\langle\bar\varphi\rangle_{\GZ}-\gamma^2\delta X$, and substitution in the transformed BRST variation yields
\begin{equation}
\bigl\langle s\bar\omega_\mu^{ab}(x)\bigr\rangle
=\bigl\langle\bar\varphi_\mu^{\prime ab}(x)\bigr\rangle+\gamma^2\delta^{ab}X_\mu(x)
=\bigl\langle\bar\varphi_\mu^{ab}(x)\bigr\rangle_{\GZ}
\qquad\text{at every finite }\Lambda .
\label{eq:ward-exact}
\end{equation}
The untranslated expectation value has no profile dependence. In an infinite-volume state invariant under translations and $O(4)$, it vanishes because a constant vector expectation value is excluded by rotational symmetry. On the Dirichlet ball, the covariant form $\delta^{ab}x_\mu h(x^2)$ is allowed, so symmetry alone does not fix the one-point function. For a rescaled profile $u\mapsto\kappa u$, the change-of-variables identity requires the contour and domination hypotheses at that value of $\kappa$.

\section{Critical spectrum and source visibility}
\label{sec:critical}

The reference for the physical interpretation is a compatible transverse truncation, where $\mathcal M_N$ is self-adjoint and a critical eigenspace is dark when it is orthogonal to the horizon source. In a noncompatible truncation, the symmetric part can lose positivity while the full compressed operator remains invertible. The latter effect is shielding caused by the compression defect and does not represent a zero mode of the self-adjoint Landau-gauge operator. The block-matrix families below illustrate this possibility algebraically, without asserting that every choice of blocks is realized by a gauge background.

\paragraph{Compression effects.}

For a background-dependent Zwanziger source $m=\bar m$, the horizon pairing is
\begin{equation}
\gamma^4\Hor(A)=m^{\mathsf T}\mathcal M_N[A]^{-1}\bar m
\label{eq:horizon-functional}
\end{equation}
Its critical behavior depends on the symmetric part, the full compressed operator, and the source projection. The quantities
\begin{equation}
\lambda_1^{\rm sym}(A)\equiv\lambda_{\min}\bigl(\Msym_N[A]\bigr),
\qquad
\operatorname{spec}\bigl(\mathcal M_N[A]\bigr),
\qquad
s_{\min}\bigl(\mathcal M_N[A]\bigr)
\label{eq:three-quantities}
\end{equation}
therefore have distinct roles. At a finite cutoff, a zero eigenvalue of $\mathcal M_N$ is equivalent to a zero smallest singular value, but away from singularity the eigenvalue distances need not measure the inverse norm, which is $s_{\min}(\mathcal M_N)^{-1}$. The boundary of the coercive domain is reached through
\begin{equation}
\lambda_1^{\rm sym}(A)\longrightarrow0 .
\label{eq:critical-limit}
\end{equation}
This need not be a spectral crossing of the full operator when the antisymmetric defect couples the critical direction to the regular subspace.

Consider a continuous family $A(\tau)$, $\tau>0$, satisfying
\begin{equation}
\lambda_1^{\rm sym}(\tau)>0\ \ (\tau>0),
\qquad
\lambda_1^{\rm sym}(\tau)\longrightarrow0^+ ,
\qquad
\liminf_{\tau\to0^+}\lambda_{r+1}^{\rm sym}(\tau)=\lambda_\star>0 ,
\label{eq:crossing-family}
\end{equation}
where $r$ eigenvalues of $\Msym_N$ reach zero and the remaining spectrum has a positive limiting gap. Write $P_c$ for the critical projector at $\tau=0$ and $m_c=\lim_{\tau\to0^+}m(A(\tau))$. The notation $\lambda_1$ denotes $\lambda_1^{\rm sym}$ when no distinction from the spectrum of $\mathcal M_N$ is needed. For a simple crossing, the coupling to the regular complement has an explicit scalar Schur denominator.

\begin{theorem}[Truncation-induced shielding at a simple crossing]
\label{thm:shielding}
In an orthonormal basis adapted to the critical direction $\psi_c$, let
\begin{equation}
\Msym(\lambda)=\begin{pmatrix}\lambda&0\\0&\Msym_\perp\end{pmatrix},
\qquad
\Masym=\begin{pmatrix}0&-a^{\mathsf T}\\a&\Masym_\perp\end{pmatrix},
\qquad
\mathcal M(\lambda)=\Msym(\lambda)+\Masym ,
\label{eq:blocks-shielding}
\end{equation}
where $\Msym_\perp\geq\sigma\mathbbm1>0$ and $a=\Masym\psi_c$. Keep $a$, $\Msym_\perp$, and $\Masym_\perp$ fixed, with $\lambda>0$ the only varying parameter. Then $D=\Msym_\perp+\Masym_\perp$ is invertible, with
\begin{equation}
\tfrac12\bigl(D^{-1}+D^{-\mathsf T}\bigr)=D^{-1}\Msym_\perp D^{-\mathsf T}>0 ,
\label{eq:sym-Dinv}
\end{equation}
and the Schur complement is
\begin{equation}
\mathfrak s(\lambda)=\lambda+\kappa_a ,
\qquad
\kappa_a\equiv a^{\mathsf T}D^{-1}a=a^{\mathsf T}\Bigl[\tfrac12\bigl(D^{-1}+D^{-\mathsf T}\bigr)\Bigr]a ,
\label{eq:schur-scalar}
\end{equation}
where $\kappa_a>0$ exactly when $a\neq0$. For $a\neq0$, the full inverse has a finite limit as $\lambda\to0^+$ and every fixed source $m=(m_c,m_\perp)$, with $m_c=\langle\psi_c,m\rangle$, satisfies
\begin{equation}
m^{\mathsf T}\mathcal M(\lambda)^{-1}m=\frac{|m_c|^2}{\lambda+\kappa_a}+O(1)
\label{eq:shielded-horizon}
\end{equation}
with a remainder bounded uniformly on $[0,\lambda_0]$. If $a=0$, the exact pairing is
\begin{equation}
m^{\mathsf T}\mathcal M(\lambda)^{-1}m=\frac{|m_c|^2}{\lambda}+m_\perp^{\mathsf T}D^{-1}m_\perp .
\label{eq:transparent-pole}
\end{equation}
Thus, for fixed antisymmetric data, a pole occurs precisely when the critical direction is uncoupled from the regular complement and has nonzero source component.
\end{theorem}

\begin{proof}
Positivity of the symmetric part gives $\ker D=\{0\}$. Multiplication of $D+D^{\mathsf T}=2\Msym_\perp$ by $D^{-1}$ and $D^{-\mathsf T}$ proves \eqref{eq:sym-Dinv}. The Schur complement is $\lambda-(-a^{\mathsf T})D^{-1}a=\lambda+\kappa_a$, and the scalar form depends only on the positive symmetric part of $D^{-1}$. Block inversion gives
\begin{equation}
\mathcal M(\lambda)^{-1}=
\begin{pmatrix}
\mathfrak s^{-1} & \mathfrak s^{-1}a^{\mathsf T}D^{-1}\\[2pt]
-D^{-1}a\,\mathfrak s^{-1} & D^{-1}-\mathfrak s^{-1}D^{-1}a\,a^{\mathsf T}D^{-1}
\end{pmatrix}.
\label{eq:block-inverse}
\end{equation}
For $a\neq0$, every entry depends on $\lambda$ through $(\lambda+\kappa_a)^{-1}\leq\kappa_a^{-1}$ and therefore has a finite limit. Contraction with $m$ isolates the displayed critical term and a bounded remainder. When $a=0$, the off-diagonal blocks vanish and the inverse is $\operatorname{diag}(\lambda^{-1},D^{-1})$.
\end{proof}

Along a crossing family, $a$ and $\lambda_1$ vary together with the background. Assume that the symmetric part of $D(\tau)$ remains uniformly positive and bounded near zero, the source and block data are continuous, and $m_c(\tau)\to m_c(0)\neq0$. With $a(\tau)=\Masym_N[A(\tau)]\psi_1(\tau)$ and $\kappa_a(\tau)=a^{\mathsf T}D^{-1}a$, define
\begin{equation}
q_-(\tau)=m_c-m_\perp^{\mathsf T}D^{-1}a,
\qquad
q_+(\tau)=m_c+a^{\mathsf T}D^{-1}m_\perp .
\label{eq:qpm}
\end{equation}
Contraction of the block inverse gives the exact expression
\begin{equation}
\gamma^4\Hor\bigl(A(\tau)\bigr)
=\frac{q_-(\tau)q_+(\tau)}{\lambda_1(\tau)+\kappa_a(\tau)}
+m_\perp^{\mathsf T}D^{-1}m_\perp,
\label{eq:shielding-rate}
\end{equation}
A uniformly positive symmetric part and boundedness of $D$ imply that $\kappa_a\to0$ exactly when $a\to0$. Consequently, if
\begin{equation}
\frac{\kappa_a(\tau)}{\lambda_1(\tau)}\longrightarrow\rho\in[0,\infty),
\label{eq:shielding-ratio}
\end{equation}
then $q_-q_+\to|m_c(0)|^2$ and
\begin{equation}
\lambda_1(\tau)\,\gamma^4\Hor\bigl(A(\tau)\bigr)
\longrightarrow\frac{|m_c(0)|^2}{1+\rho} .
\label{eq:shielding-residue}
\end{equation}
The case $\rho=0$ retains the transparent residue, while a finite positive $\rho$ reduces it. If $\kappa_a/\lambda_1\to\infty$, the $1/\lambda_1$ residue vanishes. Boundedness then requires $\liminf\kappa_a>0$, since $\kappa_a\to0$ also forces $a\to0$ and leaves the nonzero numerator $|m_c(0)|^2$. In that latter case the response grows as $|m_c(0)|^2/\kappa_a$, more slowly than $1/\lambda_1$ but without saturating.

For example, $a(\tau)=\tau a_0$ with fixed $D$ gives $\kappa_a\propto\tau^2$. The choices $\lambda_1=\tau$, $\tau^2$, and $\tau^4$ respectively preserve the residue, reduce it by a finite factor, and soften the growth from $\tau^{-4}$ to $\tau^{-2}$. Nonzero coupling at every positive $\tau$ is therefore insufficient for saturation. The ratio $\kappa_a/\lambda_1$ controls the change in critical scaling, while the limiting size of $\kappa_a$ controls boundedness.

The Schur complement of the full operator should also be distinguished from a single entry of the positive Gaussian kernel $\Bmat$. In the same basis,
\begin{equation}
\Bmat_{cc}=\psi_c^{\mathsf T}\Bmat\psi_c=\lambda+a^{\mathsf T}\Msym_\perp^{-1}a ,
\label{eq:Bcc}
\end{equation}
which agrees with $\lambda+\kappa_a$ when $\Masym_\perp=0$. For $a\neq0$, however, the block $\Bmat_{\perp\perp}=\Msym_\perp+\Masym_\perp\Msym_\perp^{-1}\Masym_\perp^{\mathsf T}+aa^{\mathsf T}/\lambda$ is singular in the limit. The entry $\Bmat_{cc}$ does not by itself determine $(\Bmat^{-1})_{cc}$, although the full scalar inverse forms agree by \eqref{eq:symmetric-inverse-identity}.

A critical cluster can contain both shielded and unshielded directions. An analytic transport of the cluster separates their contributions while allowing the regular and shielded blocks to remain non-normal.

\begin{proposition}[Analytic Laurent splitting]
\label{prop:splitting}
Keep the geometry, $(R,N,\Lambda)$, and the real coefficient-space pairings fixed on $I=(-\varepsilon,\varepsilon)$. Let $\mathcal M_N(\tau)$ and $\Msym_N(\tau)=\operatorname{sym}\mathcal M_N(\tau)$ be real analytic, with an isolated degenerate critical eigenspace at zero. Let $U(\tau)$ be analytic and orthogonal, with $U(0)=I$ and moving projector $P(\tau)=U(\tau)P_cU(\tau)^*$. Relative to $\operatorname{ran}P_c\oplus\operatorname{ran}Q_c$, $Q_c=I-P_c$, write
\begin{equation}
\widehat{\mathcal M}(\tau)=U(\tau)^*\mathcal M_N(\tau)U(\tau)
=\begin{pmatrix}A(\tau)&B(\tau)\\ C(\tau)&D(\tau)\end{pmatrix},
\qquad
\mathcal M_{\rm eff}=A-BD^{-1}C .
\label{eq:transported-blocks}
\end{equation}
Assume that all four blocks are analytic, $D(\tau)$ is invertible with $\operatorname{sym}D(\tau)\succeq\lambda_\star I$ uniformly near zero, and
\begin{align}
\widehat S_{cc}(\tau)&=\tau\Gamma+O(\tau^2),\qquad \Gamma>0,
\label{eq:transported-slope}\\
\mathcal M_{\rm eff}(\tau)&=M_0+\tau G+O(\tau^2).
\label{eq:effective-critical}
\end{align}
Suppose also that
\begin{equation}
\operatorname{ran}P_c=\mathcal H_{\rm unsh}\oplus\mathcal H_{\rm sh},
\qquad
M_0\Pi_u=\Pi_uM_0=0,
\qquad
M_s:=M_0|_{\mathcal H_{\rm sh}}\ \text{is invertible},
\label{eq:zero-order-kernel}
\end{equation}
where $\Pi_u$ is the orthogonal projector onto $\mathcal H_{\rm unsh}$, and that
\begin{equation}
G_u:=\Pi_uG\Pi_u\ \text{is invertible},
\qquad
\operatorname{sym}G_u=\Pi_u\Gamma\Pi_u>0,
\qquad
C(0)\Pi_u=0,
\quad
\Pi_uB(0)=0 .
\label{eq:reduced-gamma}
\end{equation}
Then $\mathcal M_N(\tau)$ is invertible for sufficiently small $\tau>0$ and
\begin{equation}
\mathcal M_N(\tau)^{-1}
=\frac1\tau U(\tau)\Pi_uG_u^{-1}\Pi_uU(\tau)^*+R(\tau),
\qquad
\sup_{0<\tau<\varepsilon}\|R(\tau)\|<\infty .
\label{eq:laurent-unshielded}
\end{equation}
No symmetry or normality of $M_s$ is required.
\end{proposition}

\begin{proof}
The vanishing unshielded row and column of $M_0$ give
\begin{equation}
\mathcal M_{\rm eff}(\tau)=
\begin{pmatrix}
\tau G_u+O(\tau^2)&\tau G_{us}+O(\tau^2)\\
\tau G_{su}+O(\tau^2)&M_s+\tau G_{ss}+O(\tau^2)
\end{pmatrix}.
\label{eq:inner-laurent-blocks}
\end{equation}
The shielded block $D_s(\tau)$ has a uniformly bounded analytic inverse. Its unshielded Schur complement is
\begin{equation}
K_u(\tau)=\tau G_u+O(\tau^2)
-(\tau G_{us}+O(\tau^2))D_s(\tau)^{-1}(\tau G_{su}+O(\tau^2))
=\tau\bigl(G_u+O(\tau)\bigr),
\label{eq:inner-schur}
\end{equation}
so $K_u^{-1}=\tau^{-1}G_u^{-1}+O(1)$. The block formula
\begin{equation}
\mathcal M_{\rm eff}^{-1}=
\begin{pmatrix}
K_u^{-1}&-K_u^{-1}M_{us}D_s^{-1}\\
-D_s^{-1}M_{su}K_u^{-1}&D_s^{-1}+D_s^{-1}M_{su}K_u^{-1}M_{us}D_s^{-1}
\end{pmatrix}
\label{eq:inner-block-inverse}
\end{equation}
therefore gives $\mathcal M_{\rm eff}^{-1}=\tau^{-1}\Pi_uG_u^{-1}\Pi_u+E_{\rm eff}(\tau)$, with $E_{\rm eff}$ bounded. Each off-diagonal term has order $O(\tau^{-1})O(\tau)=O(1)$, and the correction to the shielded inverse is $O(\tau)$. Lifting through the regular complement gives
\begin{equation}
\widehat{\mathcal M}^{-1}
=\binom{I}{-D^{-1}C}\mathcal M_{\rm eff}^{-1}
\begin{pmatrix}I&-BD^{-1}\end{pmatrix}
+\begin{pmatrix}0&0\\0&D^{-1}\end{pmatrix}.
\label{eq:outer-block-inverse}
\end{equation}
Analyticity and \eqref{eq:reduced-gamma} imply $C(\tau)\Pi_u=O(\tau)$ and $\Pi_uB(\tau)=O(\tau)$. The lifted off-diagonal singular terms are thus bounded, and the regular-regular term is $O(\tau)$. Conjugation by $U(\tau)$ proves the stated Laurent expansion.
\end{proof}

\paragraph{Compatible truncations and source visibility.}
The compatible realization specified in Section~\ref{sec:setting} removes the antisymmetric defect on its Landau fiber, as follows from the next proposition.

\begin{proposition}[Exact transversality in a compatible truncation]
\label{prop:compatible}
Suppose the vector and scalar coefficient spaces satisfy
\begin{equation}
\operatorname{div}\mathcal V_N\subseteq\mathcal H_N .
\label{eq:compatibility}
\end{equation}
Then for every $A_N\in\mathcal V_N$ the projected Landau condition is equivalent to pointwise transversality,
\begin{equation}
P_N(\partial_\mu A_{N,\mu})=0
\quad\Longleftrightarrow\quad
\partial_\mu A_{N,\mu}=0 ,
\label{eq:exact-transversality}
\end{equation}
and on that transverse subclass the adjoint defect vanishes identically,
\begin{equation}
\Masym_N[A_N]=0,
\qquad
\mathcal M_N[A_N]=\Msym_N[A_N]=\mathcal M_N[A_N]^{\mathsf T} .
\label{eq:compatible-symmetric}
\end{equation}
\end{proposition}

\begin{proof}
Condition~\eqref{eq:compatibility} places $\partial_\mu A_{N,\mu}$ in $\mathcal H_N$, where $P_N$ acts as the identity, which is~\eqref{eq:exact-transversality}. Substituting $\partial_\mu A_{N,\mu}=0$ into~\eqref{eq:antisymmetric-part} annihilates the right-hand side for all $v,w\in\mathcal H_N$.
\end{proof}

\begin{corollary}[Recovery of the standard Faddeev–Popov spectrum]
\label{cor:fidelity}
On the transverse subclass of Proposition~\ref{prop:compatible}, at a crossing satisfying the hypotheses of Proposition~\ref{prop:splitting}, the shielded subspace is empty,
\begin{equation}
\mathcal H_{\rm sh}=\{0\},
\qquad
\mathcal H_{\rm unsh}=\operatorname{ran}P_c ,
\qquad
G_u=\Gamma ,
\label{eq:fidelity}
\end{equation}
and the compressed operator is self-adjoint, so that
\begin{equation}
s_{\min}(\mathcal M_N)=\min_j|\lambda_j(\mathcal M_N)| .
\label{eq:selfadjoint-singular-value}
\end{equation}
On the nonnegative side of an isolated crossing the regular gap makes $s_{\min}=\lambda_{\min}$ for sufficiently small positive $\tau$.
\end{corollary}

Both statements follow from the invariance of the critical and regular subspaces in the transported spectral frame, where the off-diagonal blocks vanish and the critical block is $\tau\Gamma+O(\tau^2)$. They are algebraic and hold at fixed cutoff. Compatibility is neither a convergence nor a geometry-transfer statement.

Discrete de Rham complexes provide compatible spaces through finite-element subcomplexes, bounded cochain projections, and stability estimates~\cite{ArnoldFalkWinther2010}. Constraint-preserving schemes of this type have also been developed for the Yang-Mills equations~\cite{DroniouYangMills2022}. Another construction is $\mathcal V_N=\mathcal V_N^T\oplus\nabla\mathcal H_N$, with divergence-free $\mathcal V_N^T$ and $\partial^2\mathcal H_N\subseteq\mathcal H_N$, with the matched boundary traces explicitly realized in Section~\ref{sec:setting}. On a torus, nonzero Fourier modes with $|k|\leq\Lambda_{\rm UV}$ and all vector polarizations on the same mode set make the divergence surjective onto the scalar space. The Landau constraint selects $k_\mu\varepsilon_\mu=0$, and a smooth profile compactly supported inside the periodic cell extends periodically. This gives a different boundary-value problem from the Dirichlet ball. In either realization, divergence compatibility addresses transversality but does not close the nonlinear Zwanziger or BRST products on the retained modes, and hence supplies no finite-$N$ Slavnov-Taylor identity.

Divergence compatibility does not determine the source overlap. Because the Zwanziger source carries the labels $(\mu,c)$, its projection onto a degenerate critical space is described by the map $T_A:\mathcal K_{\rm src}\to\mathcal H_N$,
\begin{equation}
\bigl(T_Ae_{\mu c}\bigr)^a(x)=\gamma^2g\,P_N\!\left(f^{a\ell c}A_\mu^\ell\right)(x),
\label{eq:source-map}
\end{equation}
whose columns are the projected sources $\bar m_\mu^{\,\cdot\,c}$. With $T_A=P_NT_A^{\rm cont}$, the horizon functional is
\begin{equation}
\gamma^4\Hor(A)
=\operatorname{Tr}_{\mathcal K_{\rm src}}\bigl(T_A^{*}\mathcal M_N[A]^{-1}T_A\bigr)
=\operatorname{Tr}_{\mathcal H_N}\bigl(\mathcal M_N[A]^{-1}V_A\bigr),
\qquad
V_A\equiv T_AT_A^{*}\geq0 .
\label{eq:visibility-operator}
\end{equation}
The positive operator $V_A$ assigns a source weight to each subspace of ghost space, including the unshielded subspace in the Laurent expansion.

In the self-adjoint case, the overlap criterion and rank classification below are those of Proposition~2.2 of Ref.~\cite{TedescoDarkCrossings}. Here the criterion acts on the unshielded subspace, and the analytic block and source hypotheses control the remainder even for a non-self-adjoint compression.

\begin{theorem}[Visibility of the unshielded critical subspace]
\label{thm:visibility}
Let $A(\tau)$ satisfy the fixed-regulator analytic hypotheses of Proposition~\ref{prop:splitting}, and suppose that the transported source obeys
\begin{equation}
\widehat T(\tau):=U(\tau)^*T_{A(\tau)}=T_c+O(\tau)
\quad\text{in operator norm}.
\label{eq:source-rate}
\end{equation}
Since $U(0)=I$, $T_c=T_{A_c}$. Define
\begin{equation}
V_c:=P_cT_cT_c^*P_c,
\qquad
V_u:=\Pi_uV_c\Pi_u=(T_c^*\Pi_u)^*(T_c^*\Pi_u).
\label{eq:crossing-visibility}
\end{equation}
Then
\begin{equation}
\gamma^4\Hor\bigl(A(\tau)\bigr)
=\frac1\tau\operatorname{Tr}\bigl(G_u^{-1}V_u\bigr)+O(1),
\label{eq:visibility-residue}
\end{equation}
with nonnegative residue satisfying
\begin{equation}
\operatorname{Tr}\bigl(G_u^{-1}V_u\bigr)>0
\quad\Longleftrightarrow\quad T_c^*\Pi_u\neq0 .
\label{eq:visibility-criterion}
\end{equation}
If $T_c^*\Pi_u=0$, the explicit Laurent term contributes $O(\tau)$ and the full response remains bounded. In a compatible transverse truncation, $\Pi_u=P_c$ and $G_u=\Gamma>0$, so
\begin{equation}
\operatorname{rank}V_c=0:\ \text{dark};
\qquad
0<\operatorname{rank}V_c<\dim P_c:\ \text{partially visible};
\qquad
\operatorname{rank}V_c=\dim P_c:\ \text{fully visible}.
\label{eq:crossing-classification}
\end{equation}
For a rank-one compatible crossing with $\lambda_1^{\rm sym}(\tau)=\kappa\tau+O(\tau^2)$ and $\kappa>0$,
\begin{equation}
\lambda_1^{\rm sym}(\tau)\,\gamma^4\Hor\bigl(A(\tau)\bigr)\longrightarrow
C\equiv\langle\psi_c,V_{A_c}\psi_c\rangle
=\sum_{\mu,c}\bigl|\bigl\langle\psi_c,\gamma^2gf^{a\ell c}A_{c,\mu}^\ell\bigr\rangle\bigr|^2 .
\label{eq:simple-residue}
\end{equation}
\end{theorem}

\begin{proof}
The source rate gives $\widehat T(\tau)\widehat T(\tau)^*=T_cT_c^*+O(\tau)$. Inserting this expression and the Laurent expansion into the trace formula yields \eqref{eq:visibility-residue}, since the first source correction contributes $O(1)$ and the regular resolvent is bounded. Moreover,
\begin{equation}
\operatorname{sym}(G_u^{-1})
=G_u^{-\mathsf T}\operatorname{sym}(G_u)G_u^{-1}>0.
\label{eq:inverse-slope-positive}
\end{equation}
The antisymmetric part of $G_u^{-1}$ has zero trace against $V_u$, so the residue is positive exactly when $V_u\neq0$, or equivalently $T_c^*\Pi_u\neq0$. If this projection vanishes, $\widehat T(\tau)^*\Pi_u=O(\tau)$ and
\begin{equation}
\Pi_u\widehat T(\tau)\widehat T(\tau)^*\Pi_u=O(\tau^2),
\label{eq:dark-unshielded-rate}
\end{equation}
which gives an $O(\tau)$ contribution after multiplication by the singular inverse. In the compatible case, the rank classification follows from $V_c=(T_c^*P_c)^*(T_c^*P_c)$. For a simple crossing the spectral decomposition is
\begin{equation}
\gamma^4\Hor=\frac{\langle\psi_1(\tau),V_{A(\tau)}\psi_1(\tau)\rangle}{\lambda_1^{\rm sym}(\tau)}
+\sum_{j\geq2}\frac{\langle\psi_j(\tau),V_{A(\tau)}\psi_j(\tau)\rangle}{\lambda_j^{\rm sym}(\tau)} .
\label{eq:spectral-horizon}
\end{equation}
The regular sum is bounded by $\operatorname{Tr}V_A/\lambda_\star$. Multiplication by $\lambda_1^{\rm sym}$ and continuity of the critical eigenvector and source prove \eqref{eq:simple-residue}.
\end{proof}

Boundedness at a dark endpoint depends on the rate at which the source projection vanishes. In a compatible transverse truncation, assume a positive regular gap $\lambda_\star$, $\mathcal M_N(\tau)\succeq0$ for $\tau>0$, and a finite critical cluster whose eigenvalues satisfy
\begin{equation}
0<c_-\tau\leq\lambda_j(\tau)\leq c_+\tau
\quad (j\ \text{critical}).
\label{eq:critical-linear-band}
\end{equation}
With $P(\tau)$ its moving projector, the conditions
\begin{equation}
\|T_{A(\tau)}^*P(\tau)\|_{\rm HS}^2
=O\!\left(\lambda_{\min}^{\rm crit}(\tau)\right),
\qquad
\sup_{0<\tau<\varepsilon}\|T_{A(\tau)}\|_{\rm HS}<\infty,
\label{eq:source-rate-boundedness}
\end{equation}
are sufficient for a bounded horizon response. Positivity gives
\begin{equation}
0\leq
\operatorname{Tr}\!\left(\mathcal M_N(\tau)^{-1}P(\tau)V_{A(\tau)}P(\tau)\right)
\leq
\frac{\|T_{A(\tau)}^*P(\tau)\|_{\rm HS}^2}
{\lambda_{\min}^{\rm crit}(\tau)},
\label{eq:critical-source-bound}
\end{equation}
while the regular contribution is at most $\lambda_\star^{-1}\|T_{A(\tau)}^*(I-P(\tau))\|_{\rm HS}^2$. If both the source and projector are analytic and $T_c^*P_c=0$, their product is $O(\tau)$, whose squared Hilbert-Schmidt norm is $O(\tau^2)=o(\lambda_{\min}^{\rm crit})$ under the linear-band assumption.

The endpoint projection alone does not supply this rate. A critical eigenvalue $\lambda_c(\tau)=\tau$ and squared source overlap of order $\tau^{1/2}$ give a response of order $\tau^{-1/2}$. Likewise, $\mathcal M(\tau)=\operatorname{diag}(\tau^4,1)$ and $v(\tau)=\tau e_1+e_2$ are dark at the endpoint but have response $\tau^{-2}+1$. The pole residue is naturally expressed in the eigenvalue. At a transversal crossing, $\lambda_1^{\rm sym}\sim\tau\dot\lambda_1(0)$ converts it to a $1/\tau$ law. At a tangential approach with $\lambda_1^{\rm sym}\sim c\tau^2$ and nonzero limiting weight $C$, the response instead scales as $\tau^{-2}$. Vanishing of $\lambda_1^{\rm sym}\Hor$ therefore does not establish boundedness when $C=0$.

A partially visible critical space carries a pole even though some of its directions do not contribute to the residue. This dependence on eigenvector projections is already present in bounds for the ghost propagator~\cite{CucchieriMendesGhost2008,CucchieriMendes2013} and in spectral decompositions of the horizon function, which need not diverge at every point of the first horizon~\cite{CucchieriFMR1998}. Copy selection by the lowest Faddeev-Popov eigenvalue also changes infrared ghost and gluon propagators statistically~\cite{SternbeckFP2006,SternbeckMuller2013}. Overlap-dependent blindness appears in a Lorentzian source-sandwiched inverse~\cite{GuimaraesLorentzian2026}, while perturbative studies describe low eigenvalues near the first horizon~\cite{Greensite2010}. In the compressed Euclidean problem, $\kappa_a$ and $\Pi_u$ first determine which critical directions survive in the full inverse, and $V_A$ then fixes their source weight.

The cancellation of fluctuation determinants does not remove the need for a coercive bosonic integral. Restricting the quartet to one critical direction, with $\lambda>0$ and real source component $\mu$, gives
\begin{equation}
Z_{\rm crit}(\lambda,\mu)
=\int\d P\,\d Q\,\d\bar\omega\,\d\omega\;
\exp\!\left[-\frac{\lambda}{2}\bigl(P^2+Q^2\bigr)-i\sqrt2\,\mu Q+\lambda\bar\omega\omega\right]
=2\pi\,e^{-\mu^2/\lambda},
\label{eq:Zcrit-value}
\end{equation}
up to the fixed normalization of the measure, and hence
\begin{equation}
\W_{\rm crit}(\lambda,\mu)=-\ln Z_{\rm crit}=\frac{\mu^2}{\lambda}-\ln2\pi .
\label{eq:Wcrit}
\end{equation}
The $P$ and $Q$ integrals contribute $\sqrt{2\pi/\lambda}$ and $\sqrt{2\pi/\lambda}\,e^{-\mu^2/\lambda}$, respectively, while the Berezin integral contributes $\lambda$. Their product cancels the determinant exactly at every positive $\lambda$. For $\mu=0$, both $Z_{\rm crit}$ and $\W_{\rm crit}$ have finite limits, but these limits do not define an ordinary Gaussian integral at $\lambda=0$. There the bosonic integral diverges and the Berezin integral is zero, so multiplication of separately evaluated endpoint factors is undefined. For $\mu\neq0$, $Z_{\rm crit}\to0$ and $\W_{\rm crit}\sim\mu^2/\lambda$. In particular,
\begin{equation}
\lim_{\lambda\downarrow0}\lambda\W_{\rm crit}(\lambda,\mu)=\mu^2,
\label{eq:sharpness-visibility-link}
\end{equation}
for $\mu=\langle\psi_c,m\rangle$. This one-mode pole contains no coupling to a regular complement and is the unshielded case of the full Schur analysis.

\section{Radial bounds and source darkness}
\label{subsec:radial-fredholm}

Radial convexity controls the closing gap through the affine dependence of the Faddeev-Popov operator on the background. This part of the geometric bounds of Cucchieri and Mendes~\cite{CucchieriMendes2013,DudalBounds2015} has a finite-cutoff counterpart, whereas a bound on the normalized horizon function requires additional large-volume spectral information. Define
\begin{equation}
L_N\equiv\mathcal M_N[0]=P_N(-\partial^2)P_N,
\qquad
\ell_N\equiv\lambda_{\min}(L_N)>0,
\label{eq:free-dirichlet-gap}
\end{equation}
and choose a boundary background $B$ satisfying
\begin{equation}
F_N[B]=0,
\qquad
\Msym_N[B]\succeq0,
\qquad
\lambda_{\min}\!\left(\Msym_N[B]\right)=0.
\label{eq:radial-boundary-background}
\end{equation}
Along $A_\rho=\rho B$, with $0\leq\rho<1$, affine dependence gives $\mathcal M_N[\rho B]=(1-\rho)L_N+\rho\mathcal M_N[B]$. Taking symmetric parts yields
\begin{equation}
\Msym_N[A_\rho]
=(1-\rho)L_N+\rho\Msym_N[B]
\succeq(1-\rho)L_N
\succeq(1-\rho)\ell_N\mathbbm1.
\label{eq:radial-coercivity}
\end{equation}
For every coefficient vector $x$,
\begin{equation}
\|x\|\,\|\mathcal M_Nx\|
\geq \operatorname{Re}\langle x,\mathcal M_Nx\rangle
=\langle x,\Msym_Nx\rangle
\geq(1-\rho)\ell_N\|x\|^2,
\end{equation}
so
\begin{equation}
s_{\min}\!\left(\mathcal M_N[A_\rho]\right)
\geq(1-\rho)\ell_N,
\qquad
\left\|\mathcal M_N[A_\rho]^{-1}\right\|
\leq\frac{1}{(1-\rho)\ell_N}.
\label{eq:radial-singular-bound}
\end{equation}
Since $T_{A_\rho}=\rho T_B$ and $V_{A_\rho}=\rho^2V_B$, the trace pairing also satisfies
\begin{equation}
\left|\gamma^4\Hor(A_\rho)\right|
\leq
\frac{\rho^2}{(1-\rho)\ell_N}\operatorname{Tr}V_B.
\label{eq:radial-horizon-bound}
\end{equation}
These bounds apply to the full compressed inverse even when the projected Landau condition leaves an antisymmetric defect.

In a compatible transverse truncation, self-adjointness strengthens the inverse bound to
\begin{equation}
0\prec\mathcal M_N[A_\rho]^{-1}
\preceq\frac{1}{1-\rho}L_N^{-1}.
\label{eq:radial-loewner}
\end{equation}
Let $P_c$ project onto $\ker\mathcal M_N[B]$, with the regular spectrum separated from zero. Degenerate perturbation theory gives
\begin{equation}
\mathcal M_N[A_\rho]^{-1}
=
\frac{1}{1-\rho}
P_c(P_cL_NP_c)^{-1}P_c+O(1),
\label{eq:radial-critical-resolvent}
\end{equation}
where $(P_cL_NP_c)^{-1}$ is taken only on $\operatorname{ran}P_c$. The horizon response is therefore
\begin{equation}
\gamma^4\Hor(A_\rho)
=
\frac{\operatorname{Tr}\!\left[(P_cL_NP_c)^{-1}P_cV_BP_c\right]}{1-\rho}
+O(1).
\label{eq:radial-critical-horizon}
\end{equation}
For a simple critical mode, set $a_B=\langle\psi_c,L_N\psi_c\rangle>0$ and $C_B=\langle\psi_c,V_B\psi_c\rangle$. Then
\begin{equation}
\lambda_1(\rho)=a_B(1-\rho)+O\!\left((1-\rho)^2\right),
\qquad
C_B>0\ \Longrightarrow\ 
\gamma^4\Hor(A_\rho)=\frac{C_B}{a_B(1-\rho)}+O(1).
\label{eq:radial-visible-pole}
\end{equation}
If $C_B=0$ and the moving eigenspace is analytic, its source projection is $T_{A_\rho}^*P(\rho)=O(1-\rho)$. The weight $w_1(\rho)=\langle\psi_1(\rho),V_{A_\rho}\psi_1(\rho)\rangle$ consequently obeys
\begin{equation}
w_1(\rho)=O\!\left((1-\rho)^2\right)=O\!\left(\lambda_1(\rho)^2\right).
\label{eq:radial-dark-rate}
\end{equation}
For a finite-rank analytic critical cluster, $P_cV_BP_c=0$ similarly gives the basis-independent estimate
\begin{equation}
W_c(\rho)
\equiv
\left\|T_{A_\rho}^*P(\rho)\right\|_{\rm HS}^2
=O\!\left((1-\rho)^2\right).
\label{eq:radial-dark-cluster-rate}
\end{equation}
The cluster eigenvalues are $(1-\rho)\mu_j+O((1-\rho)^2)$, with positive $\mu_j$ determined by $P_cL_NP_c$. Their total contribution to the horizon pairing is thus $O(1-\rho)$ and tends to zero.

Source darkness can also be expressed as solvability of a finite linear system at the endpoint. Under the analytic block assumptions, the left and right kernels coincide even though the full compressed operator need not be self-adjoint.

\begin{proposition}[Range criterion for source darkness]
\label{prop:fredholm-darkness}
Under the hypotheses of Proposition~\ref{prop:splitting}, let $M_c=\mathcal M_N[A_c]$ and let $\Pi_u$ project onto $\mathcal H_{\rm unsh}$. Then
\begin{equation}
\ker M_c=\ker M_c^*=\mathcal H_{\rm unsh}=\operatorname{ran}\Pi_u,
\label{eq:fredholm-kernel}
\end{equation}
and
\begin{equation}
V_u=0
\quad\Longleftrightarrow\quad
T_c^*\Pi_u=0
\quad\Longleftrightarrow\quad
\operatorname{Ran}T_c\subseteq\operatorname{Ran}M_c
\quad\Longleftrightarrow\quad
\exists\,Y_c:\ M_cY_c=T_c.
\label{eq:fredholm-equivalence}
\end{equation}
For every $Y$, $\Pi_u(T_c-M_cY)=\Pi_uT_c$, so changing the source by an element of $\operatorname{Ran}M_c$ leaves its visible component unchanged. In a compatible transverse truncation, $\Pi_u=P_c$.
\end{proposition}

\begin{proof}
Since $\Msym_N[A_c]\succeq0$, $M_cx=0$ implies $\langle x,\Msym_N[A_c]x\rangle=0$, and hence $\Msym_N[A_c]x=0$ and $\Masym_N[A_c]x=0$. Applying the same argument to the adjoint gives
\begin{equation}
\ker M_c=\ker M_c^*=\ker\Msym_N[A_c]\cap\ker\Masym_N[A_c].
\label{eq:accretive-endpoint-kernel}
\end{equation}
Elimination of $D(0)$ shows that a kernel vector has critical component $x\in\ker M_0=\mathcal H_{\rm unsh}$ and regular component $-D(0)^{-1}C(0)x=0$. This identifies the common kernel. Since $V_u=(T_c^*\Pi_u)^*(T_c^*\Pi_u)$, $V_u=0$ is equivalent to $T_c^*\Pi_u=0$. The finite-dimensional Fredholm alternative gives
\begin{equation}
\operatorname{Ran}M_c=(\ker M_c^*)^\perp=\mathcal H_{\rm unsh}^\perp,
\end{equation}
so each source column is orthogonal to the common kernel exactly when it belongs to the range of $M_c$, or equivalently when $M_cY_c=T_c$ has a solution. The source-shift identity follows from $\Pi_uM_c=0$.
\end{proof}

This range criterion concerns the Euclidean endpoint operator with the Zwanziger source prescribed as a right-hand side, in contrast with the Lorentzian boundary-value formulation of the no-pole condition~\cite{GuimaraesLorentzian2026}. It also gives a direct relation between source visibility and the admissible coordinate profile. Regard $X_N$ as the map whose $(\mu,c)$ column is $X_{N,\mu}e_c$, and define
\begin{equation}
D_{A,N}[X_N]
:=T_A+\gamma^2\mathcal M_N[A]X_N.
\label{eq:profile-range-defect}
\end{equation}
At an unshielded endpoint, $\Pi_u\mathcal M_c=0$ implies
\begin{equation}
\Pi_uD_{A_c,N}[X_N]=\Pi_uT_c
\label{eq:critical-profile-defect}
\end{equation}
for every admissible profile. Thus the unshielded projection of the range defect is exactly the visible source component.

Before projection, a pointwise transverse continuum background satisfies the formal identity
\begin{equation}
\bigl(\mathcal M[A](x_\mu e_c)\bigr)^a
=-g f^{a\ell c}A_\mu^\ell,
\qquad
T_A^{\rm cont}e_{\mu c}=-\gamma^2\mathcal M[A](x_\mu e_c).
\label{eq:formal-source-range}
\end{equation}
If $x_\mu e_c$ belonged to the operator domain, this expression would place every source column in the Faddeev-Popov range. Periodic, Dirichlet, and infinite-volume $L^2$ domains obstruct that step through their respective boundary or integrability conditions. For an admissible uncompressed profile $X$, the defect $D_{A,X}=T_A^{\rm cont}+\gamma^2\mathcal M[A]X$ is
\begin{equation}
\bigl(D_{A,X}e_{\mu c}\bigr)^a
=\gamma^2\left[
-\delta^{ac}\partial^2X_\mu
+g f^{a\ell c}A_\nu^\ell
\bigl(\delta_{\nu\mu}-\partial_\nu X_\mu\bigr)
\right].
\label{eq:continuum-profile-defect}
\end{equation}
For $X_\mu=x_\mu\chi_\Lambda$, it takes the form
\begin{equation}
\bigl(D_{A,X}e_{\mu c}\bigr)^a
=\gamma^2\left[
-\delta^{ac}\bigl(2\partial_\mu\chi_\Lambda+x_\mu\partial^2\chi_\Lambda\bigr)
+g f^{a\ell c}\left(
A_\mu^\ell(1-\chi_\Lambda)
-x_\mu A_\nu^\ell\partial_\nu\chi_\Lambda
\right)
\right],
\label{eq:compact-profile-range-defect}
\end{equation}
and vanishes on the plateau but not necessarily outside it. Projection produces a second contribution,
\begin{equation}
D_{A,N}[X_N]
=P_ND_{A,X}
+\gamma^2P_N\mathcal M[A](P_NX-X),
\label{eq:profile-projection-defect}
\end{equation}
with the first term due to the profile and the second to compression. Only their combined projection onto $\mathcal H_{\rm unsh}$ enters the endpoint source weight, and this projection is independent of the admissible choice of $X_N$.

The normalized Cucchieri-Mendes bound uses more than radial coercivity. In Eqs.~(40)-(42) of Ref.~\cite{CucchieriMendes2013}, the radial gap estimate is combined with a center-of-gravity relation for free-Laplacian multiplets at very large volume to obtain~\cite{CucchieriMendes2013,DudalBounds2015}
\begin{equation}
h_{\rm CM}(A)=\frac{H(A)}{dV(N_c^2-1)}\leq\rho(A).
\label{eq:CM-bound}
\end{equation}
Assuming an analogue for the same positive self-adjoint source pairing used here, spectral positivity would imply
\begin{equation}
w_1
\leq
\gamma^4dV(N_c^2-1)\rho\,\lambda_1,
\qquad
\frac{w_1}{\gamma^4V\lambda_1}=O(1).
\label{eq:CM-source-weight}
\end{equation}
For a degenerate eigenspace with eigenvalue $\lambda$, its basis-independent source weight
\begin{equation}
W_\lambda\equiv\operatorname{Tr}(P_\lambda V_AP_\lambda)
\end{equation}
would likewise satisfy
\begin{equation}
W_\lambda
\leq
\gamma^4dV(N_c^2-1)\rho\,\lambda.
\label{eq:CM-degenerate-weight}
\end{equation}
At fixed volume and nonzero $\gamma$, these conditional bounds force the critical source weight to vanish at least linearly in its eigenvalue. Along a thermodynamic sequence, the volume factor also varies. The radial estimate alone therefore does not establish the normalized bound for this regulator. A fixed-cutoff bound of that type on the same radial family would, by \eqref{eq:radial-critical-horizon}, require $P_cV_BP_c=0$.

For this comparison, the periodic spectral construction underlying the Cucchieri-Mendes argument removes the constant color modes and normalizes the horizon function by $dV(N_c^2-1)$. Our scalar Dirichlet space contains no constant zero mode, and the normalization would be $\gamma^{-4}\operatorname{Tr}(\mathcal M_N^{-1}V_A)/[dV(N_c^2-1)]$. Thus the large-volume multiplet relation must be established for the same operator, source, and zero-mode prescription before transferring the normalized bound. The fixed-$N,R$ limit toward a critical background used here does not interchange with a thermodynamic or ultraviolet limit.

The finite free-boundary lattice identity of Schaden and Zwanziger~\cite{SchadenZwanziger1994} uses a source $B$ with a boundary contribution, removes the constant null space, and defines its horizon function with a subtraction, $H_{\rm lat}=(B,M^{-1}B)-\sum G$. Its pointwise vanishing therefore does not identify our unsubtracted Dirichlet source pairing with zero. Their coordinate shift is allowed by free lattice boundary conditions, whereas $x_\mu$ violates the scalar Dirichlet boundary condition here. A comparison must retain those source, subtraction, and boundary terms, which can remain relevant at large volume.

A continuum realization of darkness is furnished by the transverse $SU(2)$ family $A_\mu^a=\xi\eta_{a\mu\nu}x_\nu\phi(r)$ on flat $\R^4$, with $r=|x|$ and $\xi\in\R$, where $\eta_{a\mu\nu}$ denotes the self-dual 't~Hooft symbols and $\phi\in C_c^\infty((0,\infty))$ is real. Its Faddeev-Popov operator preserves hyperspherical degree, while every Zwanziger source column has degree one, as established in Theorem~5.1 of Ref.~\cite{TedescoDarkCrossings}. Critical states of any other degree are therefore orthogonal to the source. For nonnegative smooth radial shells satisfying the normalization and width condition of Theorem~5.2 of that reference, the first loss of nonnegativity on either amplitude branch occurs at degree $n\geq2$.

For a radial mode $f(r)$, writing $u=r^{3/2}f$ gives the radial Hilbert space $L^2((0,\infty),\d r)$ and the Friedrichs realization of
\begin{equation*}
h_{n,c}(\xi)=-\frac{\d^2}{\d r^2}
+\frac{(n+\tfrac12)(n+\tfrac32)}{r^2}+2g\xi c\,\phi(r),
\end{equation*}
where $c$ is the color--angular eigenvalue, with $c=-1$ and $c=1/2$ in the source sector $n=1$. The form domain is $H_0^1(0,\infty)$, and the operator domain consists of its elements for which $h_{n,c}(\xi)u\in L^2$ distributionally. On a Dirichlet ball the form domain is $H_0^1(0,R)$, with the analogous operator-domain condition and $u(R)=0$. In full space, zero already belongs to the essential spectrum, so the positivity threshold is distinct from the isolated finite-cutoff gap closing used in the Laurent expansions. With $\rho(r)=r^{5/2}g\xi\phi(r)$, the visible inverse forms are defined by $\lim_{\varepsilon\downarrow0}\langle\rho,(h_{1,c}(\xi)+\varepsilon)^{-1}\rho\rangle$. Proposition~5.4 of Ref.~\cite{TedescoDarkCrossings} establishes their finiteness at the dark thresholds. Compact support and strict separation from the attractive degree-one threshold control the attractive visible channel, while the other visible channel is repulsive. This finiteness concerns the source inverse forms and supplies neither a bounded full-space inverse nor a uniform spectral gap.

\section{Composite source and resolvent response}
\label{sec:composite}

The dimension-two insertion of the localizing quartet probes higher powers of the inverse that enters the horizon functional. For the conventional color-singlet contraction,
\begin{equation}
\Sigma_N=\bar\varphi_i^a\varphi_i^a-\bar\omega_i^a\omega_i^a,
\qquad i=(\mu,c),
\label{eq:sigma-def}
\end{equation}
the two fields in each bilinear lie in the retained mode spaces, but their pointwise product is not projected again. It is paired with a real external source $J\in L^\infty(B_R)$ that need not belong to $\mathcal H_N$. Other independent color projections are possible~\cite{Gracey2010}, but the scalar source considered here couples to this contraction. The continuum relation $\Sigma=s(\bar\omega\varphi)$ identifies the operator in the untruncated theory, while its finite-cutoff response follows directly from the Gaussian integral.

The affine map transports the observable as well as the action. With $u_\mu^{ac}=\gamma^2\delta^{ac}X_\mu$, its pullback is
\begin{equation}
\Sigma_{\rm pb}
=(\bar\varphi'+u)(\varphi'+u)-\bar\omega\omega
=\bar\varphi'\varphi'-\bar\omega\omega
+u\varphi'+\bar\varphi'u+u^2.
\label{eq:sigma-pullback}
\end{equation}
The polynomial $\bar\varphi'\varphi'-\bar\omega\omega$ alone is a different insertion because it omits the linear and constant terms. Coupling a real scalar source through $S\mapsto S+\int J\Sigma$, and writing $J_N=P_N\mathsf M_JP_N$ with $\mathsf M_Jf=Jf$ acting trivially on internal labels, gives
\begin{equation}
\mathcal M_N(J)=\mathcal M_N-J_N,
\qquad
R_J=\mathcal M_N(J)^{-1}.
\label{eq:source-deformed-M}
\end{equation}
Equivalently, $(J_N)_{rs}=\int_{B_R}\d^4x\,\psi_r(x)J(x)\psi_s(x)$ in a real orthonormal basis. This prescription compresses the two field indices only. Projecting the bilinear itself would instead replace $J$ by $P_NJ$ and define a different insertion. In particular, an untruncated constant source $J=j$ gives $J_N=jI$, although the constant function is not a Dirichlet mode. We restrict $J$ to the open domain where $\operatorname{sym}\mathcal M_N(J)$ has a positive lower bound.

\begin{theorem}[Exact horizon response at nonzero source]
\label{thm:sigma-response}
At fixed background in the source-deformed positivity domain, integration of the localizing quartet gives, up to $J$-independent normalizations,
\begin{equation}
\W_{\rm hor}[J;A]
=m^{\mathsf T}R_J\bar m .
\label{eq:sigma-resolvent}
\end{equation}
The bosonic and Grassmann fluctuation determinants cancel for every admissible $J_N$, and coupling the source to the transported insertion~\eqref{eq:sigma-pullback} returns the same functional, with no residual dependence on $u$ or on the profile.
\end{theorem}

\begin{proof}
The two kernels carry the common deformation $K_J=K_N+J_N=-\mathcal M_N(J)$. A Berezin translation removes the one-sided Grassmann source, including the ghost-localizer vertex, without changing the determinant, while completion of the bosonic square produces the inverse determinant together with the source pairing. Their product contains no fluctuation determinant and gives~\eqref{eq:sigma-resolvent}. In the translated variables the linear sources are $K_Ju+\bar m$ and $u^{\mathsf T}K_J+m^{\mathsf T}$, whose Gaussian contraction cancels the explicit $u^{\mathsf T}K_Ju$ term, and $u^{\mathsf T}\bar m=m^{\mathsf T}u=0$ removes the mixed contractions.
\end{proof}

Variation of the inverse gives
\begin{equation}
\delta R_J=R_J\,(\delta J_N)\,R_J,
\label{eq:resolvent-variation}
\end{equation}
and hence
\begin{equation}
\delta\W_{\rm hor}
=m^{\mathsf T}R_J(\delta J_N)R_J\bar m.
\label{eq:sigma-first-variation}
\end{equation}
This is a left-right pairing through the compressed multiplication operator $\delta J_N$. For a nonsymmetric $R_J$, it need not be a norm, and positivity of the symmetric part of $\mathcal M_N(J)$ does not imply positivity of $m^{\mathsf T}R_J^2m$. At fixed background and fixed $\gamma$, a constant source $J_N=j\mathbbm1$ gives
\begin{equation}
\frac{\d\W_{\rm hor}}{\d j}
=m^{\mathsf T}(\mathcal M_N-j\mathbbm1)^{-2}\bar m.
\label{eq:sigma-constant-response}
\end{equation}
Repeated use of $\partial_jR_j=R_j^2$ yields $\partial_j^kR_j=k!R_j^{k+1}$ and
\begin{equation}
\left.\frac{\partial^k\W_{\rm hor}}{\partial j^k}\right|_{j=0}
=k!\,m^{\mathsf T}\mathcal M_N^{-(k+1)}\bar m,
\qquad k=0,1,2,\ldots ,
\label{eq:sigma-source-hierarchy}
\end{equation}
without a normality assumption. Along a critical family these derivatives are evaluated at $j=0$ before the gap is closed. In the compatible case the source-domain boundary $j=\lambda_1$ approaches zero, so no common source neighborhood extends along the entire family. For a positive self-adjoint kernel and coincident real sources,
\begin{equation}
\frac{\d\W_{\rm hor}}{\d j}
=\big\|(\mathcal M_N-j\mathbbm1)^{-1}m\big\|^2\geq0,
\label{eq:sigma-positive-compatible}
\end{equation}
and the hierarchy consists of positive higher inverse moments.

\begin{theorem}[Critical source-response hierarchy]
\label{thm:sigma-critical-response}
Keep $(R,N,\Lambda)$ and $\gamma$ fixed. Let $A(\tau)\to A_c$ approach a simple crossing in a compatible transverse truncation, with $\mathcal M_N[A(\tau)]$ positive and self-adjoint for $\tau>0$. Suppose that its normalized critical eigenvector converges to $\psi_c$, its eigenvalue $\lambda_1(\tau)$ tends to zero, and the regular eigenvalues satisfy $\lambda_\ell(\tau)\geq\lambda_\star>0$ for $\ell\geq2$. Assume $V_{A(\tau)}\to V_{A_c}$ in operator norm and
\begin{equation}
C=\langle\psi_c,V_{A_c}\psi_c\rangle>0
\label{eq:sigma-critical-C}
\end{equation}
with the rate
\begin{equation}
w_1(\tau):=\langle\psi_1(\tau),V_{A(\tau)}\psi_1(\tau)\rangle
=C+O\bigl(\lambda_1(\tau)\bigr).
\label{eq:sigma-overlap-rate}
\end{equation}
Then
\begin{align}
\W_{\rm hor}[0;A(\tau)]
&=\frac{C}{\lambda_1}+O(1),
\label{eq:sigma-critical-W}\\
\left.\frac{\partial\W_{\rm hor}}{\partial j}\right|_{j=0}
&=\frac{C}{\lambda_1^2}+O(\lambda_1^{-1}),
\label{eq:sigma-critical-first}\\
\left.\frac{\partial^2\W_{\rm hor}}{\partial j^2}\right|_{j=0}
&=\frac{2C}{\lambda_1^3}+O(\lambda_1^{-2})
\label{eq:sigma-critical-second}
\end{align}
as $\lambda_1\downarrow0$, independently of whether the eigenvalue is linear in the path parameter.
\end{theorem}

\begin{proof}
The spectral decomposition gives
\begin{equation}
\operatorname{Tr}\bigl(\mathcal M_N^{-p}V_A\bigr)
=\frac{w_1(\tau)}{\lambda_1^p}
+\sum_{\ell\geq2}\frac{\langle\psi_\ell,V_A\psi_\ell\rangle}{\lambda_\ell^p},
\qquad p=1,2,3.
\label{eq:sigma-critical-power}
\end{equation}
At fixed cutoff the regular sum is nonnegative and bounded by $\lambda_\star^{-p}\operatorname{Tr}V_A$. Substitution of the source-weight rate therefore gives $\operatorname{Tr}(\mathcal M_N^{-p}V_A)=C\lambda_1^{-p}+O(\lambda_1^{1-p})$. The full source pairing in \eqref{eq:sigma-source-hierarchy} equals this trace with $p=k+1$. Taking $k=0,1,2$ proves the three asymptotics, with the factor $2$ in the second derivative supplied by $k!$.
\end{proof}

For the pure horizon pairing, without an additive $j$-independent term, the common residue cancels in
\begin{equation}
\mathcal R_\Sigma
\equiv
\left.
\frac{\bigl(\partial_j\W_{\rm hor}\bigr)^2}
{\W_{\rm hor}\,\tfrac12\partial_j^2\W_{\rm hor}}
\right|_{j=0}
\longrightarrow1
\qquad(\lambda_1\to0^+).
\label{eq:sigma-critical-ratio}
\end{equation}
At any positive self-adjoint background with $V_A\neq0$, the same ratio satisfies
\begin{equation}
0<\mathcal R_\Sigma\leq1.
\label{eq:sigma-ratio-bound}
\end{equation}
Writing $w_\ell=\langle\psi_\ell,V_A\psi_\ell\rangle\geq0$, Cauchy-Schwarz applied to $a_\ell=\sqrt{w_\ell/\lambda_\ell}$ and $b_\ell=\sqrt{w_\ell/\lambda_\ell^3}$ gives
\begin{equation*}
\left(\sum_\ell\frac{w_\ell}{\lambda_\ell^2}\right)^2
\leq
\left(\sum_\ell\frac{w_\ell}{\lambda_\ell}\right)
\left(\sum_\ell\frac{w_\ell}{\lambda_\ell^3}\right),
\end{equation*}
with equality exactly when all eigenvalues carrying nonzero weight coincide. The ratio is dimensionless and independent of the overall source normalization. Its critical limit requires only $w_1\to C>0$ and a bounded regular contribution, while the stronger rate in \eqref{eq:sigma-overlap-rate} fixes the remainder orders. Equality is also attained by a source supported on a degenerate regular eigenspace, so it characterizes concentration at one spectral value, not criticality by itself.

At a simple dark crossing, the critical part of the $k$th derivative is $k!w_1/\lambda_1^{k+1}$. With a uniform regular gap and bounded source, positivity gives
\begin{equation}
\left.\partial_j^k\W_{\rm hor}\right|_{j=0}=O(1)
\quad\Longleftrightarrow\quad
w_1(\tau)=O\!\left(\lambda_1(\tau)^{k+1}\right),
\qquad k=0,1,2.
\label{eq:sigma-dark-rate}
\end{equation}
A bounded horizon term therefore requires less source suppression than a bounded first or second derivative. Only an exact selection rule $w_1(\tau)\equiv0$ removes the critical contribution at every order. Along an analytic radial dark branch, $\lambda_1\asymp1-\rho$ and the critical weight is $O((1-\rho)^2)$, so
\begin{equation}
\left.\partial_j^k\W_{\rm hor}^{\rm crit}\right|_{j=0}
=O\!\left((1-\rho)^{1-k}\right),
\qquad k=0,1,2.
\label{eq:radial-dark-composite-rate}
\end{equation}
The horizon contribution tends to zero and the first derivative stays bounded, while boundedness of the second derivative requires an additional power of source suppression.

For the separate translation-invariant free-space tree-level check, with $\mathcal M_0(p)=p^2$, the bare choice $j=-M^2$ gives
\begin{equation}
R_{-M^2}(p)=\frac{1}{p^2+M^2}.
\label{eq:sigma-massive-resolvent}
\end{equation}
Including the usual gluonic mass channel $m_g^2A^2/2$ and expanding the horizon pairing to quadratic order in $A$ yields
\begin{equation}
\Gamma^{(2)}_{AA}(p)
=p^2+m_g^2+\frac{2N_cg^2\gamma^4}{p^2+M^2},
\label{eq:rgz-inverse-check}
\end{equation}
and hence
\begin{equation}
\mathcal D_{\RGZ}(p^2)
=\frac{p^2+M^2}
{(p^2+m_g^2)(p^2+M^2)+2N_cg^2\gamma^4},
\label{eq:rgz-propagator-check}
\end{equation}
which is the tree-level RGZ propagator~\cite{Dudal2008,Dudal2009,Dudal2011}, with the bare scalar insertion here fixing only the auxiliary-mass denominator. A complete treatment of the condensate also requires LCO renormalization, contact counterterms, operator mixing, and the effective potential~\cite{Dudal2011,DudalVacuum2019}, so the fixed-background moment identities should be understood as relations within the bare horizon sector rather than as renormalized composite correlators.

\section{Discussion}
\label{sec:discussion}

In this work, we have used a Maggiore-Schaden-type recentering to examine the horizon term from a different angle, under the fixed-background coercivity and contour hypotheses, with gauge-field integration conditional on the stated domination assumptions. Making a coordinate shift by a smooth compact profile allows the transformation to remain inside the Dirichlet domain, while the original Maggiore-Schaden form is recovered wherever the profile coincides with $x_\mu$~\cite{MaggioreSchaden1994}. After integrating out the localizing quartet, the horizon function comes back with the same coefficient and no residual dependence on the profile, and the recentering itself therefore does not add anything to the horizon term that could be interpreted as a new vacuum contribution. What it does is separate the change of variables from the question of which vacuum is actually selected, and this distinction becomes especially relevant once boundary conditions are taken into account~\cite{DudalSorella2012,SchadenZwanziger2015a,SchadenZwanziger2015b}.

In the compatible self-adjoint setting, as the first Gribov horizon is approached, the closing Faddeev-Popov eigenvalue sets the singular scale, but the size of the horizon response still depends on how the corresponding mode overlaps with the Zwanziger source. Configurations with similarly small eigenvalues can therefore contribute very differently, as already suggested by spectral studies of the ghost propagator and of the horizon function~\cite{CucchieriMendesGhost2008,CucchieriMendes2013,CucchieriFMR1998}. The operator $V_A=T_AT_A^*$ expresses this dependence without choosing a basis in the critical space. In the transverse $SU(2)$ family, the source lies in the hyperspherical sector of degree one, so zero modes in the other angular sectors remain in the Faddeev-Popov spectrum without contributing to the horizon pairing~\cite{TedescoDarkCrossings}.

A generic Galerkin projection may leave an antisymmetric part in the regulated Faddeev-Popov operator, although this disappears when the scalar and vector spaces are chosen compatibly with the divergence. The associated shielding concerns a boundary of symmetric-part coercivity and can occur with an invertible full matrix, whereas source darkness concerns an actual critical eigenspace with zero source overlap. The existence of source-dark critical modes also tells us nothing yet about how often such configurations are actually sampled by the Yang-Mills measure. The RGZ extension has the same limited scope, where transporting $\Sigma=\bar\varphi\varphi-\bar\omega\omega$ through the recentering gives the resolvent $(\mathcal M_N-J_N)^{-1}$ and, at $J=-M^2$, reproduces the auxiliary denominator of the tree-level RGZ propagator~\cite{Dudal2008,Dudal2009,Dudal2011}. This is still a bare fixed-background result, while the condensate dynamics belongs to the full LCO treatment, including operator mixing, contact counterterms and the effective potential~\cite{Dudal2011,DudalVacuum2019}.

Exact darkness in the radial examples of Ref.~\cite{TedescoDarkCrossings} is protected by angular symmetry. For a smooth finite-volume transverse perturbation of size $\epsilon$, suppose an isolated critical subspace remains separated from its complement and acquires first-order overlap with the perturbed source. The squared weight is then $O(\epsilon^2)$, and its contribution has scale $\epsilon^2/\lambda_1$ on the positive side. Thus weak symmetry breaking can become relevant sufficiently close to the horizon. This perturbative observation does not establish stability of the full-space thresholds or their statistical weight.

The ultraviolet cutoff $N$, volume radius $R$, profile scale $\Lambda$, and coercive lower bound $\delta$ play distinct roles. The finite-dimensional identities are proved at fixed regulators under their stated hypotheses. Their integrated and source-differentiated versions remain conditional on the corresponding domination assumptions. The pointwise compact-profile limit does not commute with integration, as in \eqref{eq:vacuum-noncommutativity}, and simultaneous regulator removal would require uniform bounds not supplied here. The continuum examples of Ref.~\cite{TedescoDarkCrossings} are separate operator results, while the free-space propagators and Appendix~\ref{app:gap} are consistency checks. A continuum limit of the regulated functional integral remains open.

The next step is to carry this comparison beyond fixed cutoff. Gauge copies with similar $\lambda_1$ but different source overlaps would provide a direct test of how much the lowest eigenvalue alone misses, while the continuum problem is to control the same source projection together with the resolvent and the Yang-Mills measure as the volume grows. This is where one can decide if the source-dark directions discussed here remain exceptional or become part of the near-horizon dynamics.

\appendix

\section{Conventions and notation}
\label{app:conventions}

The Euclidean metric is $\delta_{\mu\nu}=\operatorname{diag}(1,1,1,1)$, with $px=p_\mu x_\mu$. Lowercase Latin indices label adjoint color components of $SU(N_c)$ and Greek indices label spacetime components. The structure constants are totally antisymmetric, and $N$ denotes the Galerkin cutoff. The Fourier convention $\tilde\phi(p)=\int\d^4x\,\phi(x)e^{-ipx}$ gives $\partial_\mu\phi\to ip_\mu\tilde\phi$ and $x_\mu\phi\to i\partial_{p_\mu}\tilde\phi$.

We write $Z=\int D\Phi\,e^{-S_{\rm tot}}$, $W[J]=\ln Z[J]$, and $\W[J]=-W[J]$, while $\Gamma^{(2)}_{AA}$ denotes the one-particle-irreducible gluon two-point function. The matrix $\Gamma$ in \eqref{eq:transported-slope} is the critical crossing slope. Connected second derivatives of $W$ consequently have the opposite sign to those of $\W$. The horizon contribution is denoted by $\W_{\rm hor}[J;A]$. The multi-index $i=(\mu,c)$ has $4(N_c^2-1)$ values, and pairings such as $m^{\mathsf T}\mathcal M^{-2}\bar m$ include position, color, and source-label contractions on the retained spaces. The scalar mass source is $-M^2$, while $\mathcal M=-K$ denotes the Faddeev-Popov operator.

Finite-dimensional identities use $d=4$. The one-loop calculation on the untranslated action uses $d=4-2\epsilon$ and $1/\bar\epsilon=1/\epsilon-\gamma_E+\ln4\pi$ in the $\overline{\rm MS}$ scheme.

\section{One-loop gap equation and normalization}
\label{app:gap}

This appendix checks the coefficient of the vacuum term preserved by the recentering against the known one-loop gap equation~\cite{Gracey2006}. It is a separate free-space dimensional-regularization calculation, not a new gap equation or a removal of the Dirichlet-Galerkin regulators.

The finite part of the gap equation depends on the dimensional continuation of the vacuum term. Set $\lambda^4=2N_cg^2\gamma^4$, $L=\ln(\lambda^4/\mu^4)$, and $\int_k=\mu^{2\epsilon}\int\d^dk/(2\pi)^d$, with $d=4-2\epsilon$ and $1/\bar\epsilon=1/\epsilon-\gamma_E+\ln4\pi$. The required integrals are
\begin{align}
\int_k\ln\bigl(k^2+m^2\bigr)&=-\frac{m^4}{32\pi^2}\Bigl[\frac{1}{\bar\epsilon}+\frac32-\ln\frac{m^2}{\mu^2}\Bigr]+O(\epsilon),
\label{eq:CW}\\
\int_k\ln\bigl(k^4+\lambda^4\bigr)&=\frac{\lambda^4}{16\pi^2}\Bigl[\frac{1}{\bar\epsilon}-\frac{L}{2}+\frac32\Bigr]+O(\epsilon),
\label{eq:quartic-log}\\
\int_k\frac{1}{k^4+\lambda^4}&=\frac{1}{16\pi^2}\Bigl[\frac{1}{\bar\epsilon}+1-\frac{L}{2}\Bigr]+O(\epsilon).
\label{eq:quartic-tadpole}
\end{align}
The first follows by integrating $\int_k(k^2+m^2)^{-1}=\mu^{2\epsilon}\Gamma(1-d/2)(4\pi)^{-d/2}(m^2)^{d/2-1}$ with respect to $m^2$, fixing the integration constant by the vanishing scaleless limit. Factoring $k^4+\lambda^4=(k^2+i\lambda^2)(k^2-i\lambda^2)$ gives the second integral by continuation to the conjugate masses, whose imaginary parts cancel. Partial fractions, or differentiation with respect to $\lambda^4$, give the third.

There are $d-1$ transverse polarizations per color, and the auxiliary determinants cancel pairwise. Keeping the tree-level $d$ dependence gives
\begin{equation}
\mathcal E(\gamma^4)=-d\,(N_c^2-1)\gamma^4+\frac{(d-1)(N_c^2-1)}{2}\int_k\ln\bigl(k^4+\lambda^4\bigr).
\label{eq:vacuum-energy}
\end{equation}
Stationarity of the bare expression, with $\partial\lambda^4/\partial\gamma^4=2N_cg^2$, yields
\begin{equation}
1=\frac{d-1}{d}\,\frac{N_cg^2}{16\pi^2}\Bigl[\frac{1}{\bar\epsilon}+1-\frac{L}{2}\Bigr].
\label{eq:bare-gap}
\end{equation}
Expansion of the coefficient before subtraction gives
\begin{equation}
\frac{d-1}{d}=\frac{3-2\epsilon}{4-2\epsilon}=\frac34\Bigl(1-\frac{\epsilon}{6}\Bigr)+O(\epsilon^2).
\label{eq:d-ratio}
\end{equation}
The evanescent term thus contributes $-1/6$ to the finite part, and minimal subtraction yields
\begin{equation}
1=\frac{3N_cg^2}{128\pi^2}\Bigl[\frac53-L\Bigr],
\qquad
\ln\frac{2N_cg^2\gamma_*^4}{\mu^4}=\frac53-\frac{128\pi^2}{3N_cg^2} .
\label{eq:gap-equation}
\end{equation}
Fixing the tree term at $d=4$ before this step would replace the correction $-1/6$ by $-2/3$ and the gap constant $5/3$ by $2/3$, changing $\gamma_*^4$ by a factor of $e$. The tree-level $\epsilon$ dependence must therefore be retained when relating subtraction and stationarity.

In Gracey's convention~\cite{Gracey2006}, the denominator is $p^4+C_A\gamma_G^4$, with $C_A\gamma_G^4=\lambda^4$ and $\gamma_G^4=2g^2\gamma^4$. His one-loop condition $1=C_Aa[5/8-(3/8)\ln(C_A\gamma_G^4/\mu^4)]$, where $a=g^2/(16\pi^2)$, agrees with \eqref{eq:gap-equation}. The solution $2N_cg^2\gamma_*^4=\mu^4e^{5/3}\exp[-128\pi^2/(3N_cg^2)]$ exhibits dimensional transmutation and the nonperturbative exponential dependence of Gribov's cutoff result~\cite{Gribov1978}. Its scale evolution involves the running coupling, the anomalous scaling of the renormalized Gribov parameter, and a consistently truncated effective action satisfying the renormalization identities~\cite{MaggioreSchaden1994,Zwanziger1993,CapriGZ2011,Gracey2006}.

\section*{Declarations}

\subsection*{Funding}
The author received no specific funding for this work.

\subsection*{Availability of data and material}
Not applicable. 

\subsection*{Competing interests}
The author declares that he has no competing interests.

\subsection*{Authors Contributions}
Not applicable. 

\subsection*{Use of generative AI}
Generative AI tools were used during preparation of the manuscript for editorial revision, for algebraic consistency checks, and for bibliographic retrieval. All definitions, hypotheses, statements, and proofs were formulated and verified by the author, who reviewed and edited the resulting text and assumes responsibility for the content of the publication.


\begin{thebibliography}{99}

\bibitem{Gribov1978}
V. N. Gribov, \textit{Quantization of non-Abelian gauge theories}, Nucl. Phys. B \textbf{139}, 1-19 (1978).

\bibitem{Singer1978}
I. M. Singer, \textit{Some remarks on the Gribov ambiguity}, Commun. Math. Phys. \textbf{60}, 7-12 (1978), doi:\href{https://doi.org/10.1007/BF01609471}{\nolinkurl{10.1007/BF01609471}}.

\bibitem{SemenovFranke1986}
M. A. Semenov-Tyan-Shanskii and V. A. Franke, \textit{A variational principle for the Lorentz condition and restriction of the domain of path integration in non-Abelian gauge theory}, Zap. Nauchn. Sem. LOMI \textbf{120}, 159-165 (1982); English translation: J. Sov. Math. \textbf{34}, 1999-2004 (1986), doi:\href{https://doi.org/10.1007/BF01095108}{\nolinkurl{10.1007/BF01095108}}.

\bibitem{DellAntonioZwanziger1991}
G. Dell'Antonio and D. Zwanziger, \textit{Every gauge orbit passes inside the Gribov horizon}, Commun. Math. Phys. \textbf{138}, 291-299 (1991), doi:\href{https://doi.org/10.1007/BF02099494}{\nolinkurl{10.1007/BF02099494}}.

\bibitem{vanBaal1992}
P. van Baal, \textit{More (thoughts on) Gribov copies}, Nucl. Phys. B \textbf{369}, 259-275 (1992).

\bibitem{OrlandSemenoff2000}
P. Orland and G. W. Semenoff, \textit{Extremal curves in (2+1)-dimensional Yang-Mills theory}, Nucl. Phys. B \textbf{576}, 627-654 (2000), arXiv:hep-th/9912009, doi:\href{https://doi.org/10.1016/S0550-3213(00)00134-6}{\nolinkurl{10.1016/S0550-3213(00)00134-6}}.

\bibitem{Henyey1979}
F. S. Henyey, \textit{Gribov ambiguity without topological charge}, Phys. Rev. D \textbf{20}, 1460 (1979), doi:\href{https://doi.org/10.1103/PhysRevD.20.1460}{\nolinkurl{10.1103/PhysRevD.20.1460}}.

\bibitem{GuimaraesSorella2011}
M. S. Guimaraes and S. P. Sorella, \textit{A few remarks on the zero modes of the Faddeev-Popov operator in the Landau and maximal Abelian gauges}, J. Math. Phys. \textbf{52}, 092302 (2011), arXiv:1106.3944, doi:\href{https://doi.org/10.1063/1.3641892}{\nolinkurl{10.1063/1.3641892}}.

\bibitem{CapriZeroModes2012}
M. A. L. Capri, M. S. Guimaraes, S. P. Sorella and D. G. Tedesco, \textit{A study of the zero modes of the Faddeev-Popov operator in Euclidean Yang-Mills theories in the Landau gauge in $d=2,3,4$ dimensions}, Eur. Phys. J. C \textbf{72}, 1939 (2012), arXiv:1201.2445, doi:\href{https://doi.org/10.1140/epjc/s10052-012-1939-8}{\nolinkurl{10.1140/epjc/s10052-012-1939-8}}.

\bibitem{LandimZeroModes2014}
R. R. Landim, L. C. Q. Vilar, O. S. Ventura and V. E. R. Lemes, \textit{On the zero modes of the Faddeev-Popov operator in the Landau gauge}, J. Math. Phys. \textbf{55}, 022901 (2014), arXiv:1212.4098.

\bibitem{LandimCopies2014}
R. R. Landim, V. E. R. Lemes, O. S. Ventura and L. C. Q. Vilar, \textit{Revisiting Gribov's copies inside the horizon}, Eur. Phys. J. C \textbf{74}, 3069 (2014), arXiv:1402.6235, doi:\href{https://doi.org/10.1140/epjc/s10052-014-3069-y}{\nolinkurl{10.1140/epjc/s10052-014-3069-y}}.

\bibitem{Zwanziger1989}
D. Zwanziger, \textit{Action from the Gribov horizon}, Nucl. Phys. B \textbf{321}, 591-604 (1989).

\bibitem{ZwanzigerLocal1989}
D. Zwanziger, \textit{Local and renormalizable action from the Gribov horizon}, Nucl. Phys. B \textbf{323}, 513-544 (1989).

\bibitem{Zwanziger1993}
D. Zwanziger, \textit{Renormalizability of the critical limit of lattice gauge theory by BRS invariance}, Nucl. Phys. B \textbf{399}, 477-513 (1993).

\bibitem{MaggioreSchaden1994}
N. Maggiore and M. Schaden, \textit{Landau gauge within the Gribov horizon}, Phys. Rev. D \textbf{50}, 6616-6625 (1994), arXiv:hep-th/9310111, doi:\href{https://doi.org/10.1103/PhysRevD.50.6616}{\nolinkurl{10.1103/PhysRevD.50.6616}}.

\bibitem{VandersickelZwanziger2012}
N. Vandersickel and D. Zwanziger, \textit{The Gribov problem and QCD dynamics}, Phys. Rep. \textbf{520}, 175-251 (2012), arXiv:1202.1491, doi:\href{https://doi.org/10.1016/j.physrep.2012.07.003}{\nolinkurl{10.1016/j.physrep.2012.07.003}}.

\bibitem{CapriNoPole2013}
M. A. L. Capri, D. Dudal, M. S. Guimaraes, L. F. Palhares and S. P. Sorella, \textit{An all-order proof of the equivalence between Gribov's no-pole and Zwanziger's horizon conditions}, Phys. Lett. B \textbf{719}, 448-453 (2013), arXiv:1212.2419, doi:\href{https://doi.org/10.1016/j.physletb.2013.01.039}{\nolinkurl{10.1016/j.physletb.2013.01.039}}.

\bibitem{DudalHorizon2010}
D. Dudal, S. P. Sorella and N. Vandersickel, \textit{More on the renormalization of the horizon function of the Gribov-Zwanziger action and the Kugo-Ojima Green function(s)}, Eur. Phys. J. C \textbf{68}, 283-298 (2010), arXiv:1001.3103, doi:\href{https://doi.org/10.1140/epjc/s10052-010-1304-8}{\nolinkurl{10.1140/epjc/s10052-010-1304-8}}.

\bibitem{Dudal2008}
D. Dudal, S. P. Sorella, N. Vandersickel and H. Verschelde, \textit{New features of the gluon and ghost propagator in the infrared region from the Gribov-Zwanziger approach}, Phys. Rev. D \textbf{77}, 071501(R) (2008), arXiv:0711.4496, doi:\href{https://doi.org/10.1103/PhysRevD.77.071501}{\nolinkurl{10.1103/PhysRevD.77.071501}}.

\bibitem{Dudal2009}
D. Dudal, J. A. Gracey, S. P. Sorella, N. Vandersickel and H. Verschelde, \textit{Refinement of the Gribov-Zwanziger approach in the Landau gauge: infrared propagators in harmony with the lattice results}, Phys. Rev. D \textbf{78}, 065047 (2008), arXiv:0806.4348, doi:\href{https://doi.org/10.1103/PhysRevD.78.065047}{\nolinkurl{10.1103/PhysRevD.78.065047}}.

\bibitem{Dudal2011}
D. Dudal, S. P. Sorella and N. Vandersickel, \textit{Dynamical origin of the refinement of the Gribov-Zwanziger theory}, Phys. Rev. D \textbf{84}, 065039 (2011), arXiv:1105.3371, doi:\href{https://doi.org/10.1103/PhysRevD.84.065039}{\nolinkurl{10.1103/PhysRevD.84.065039}}.

\bibitem{BaulieuSorella2009}
L. Baulieu and S. P. Sorella, \textit{Soft breaking of BRST invariance for introducing non-perturbative infrared effects in a local and renormalizable way}, Phys. Lett. B \textbf{671}, 481-485 (2009), arXiv:0808.1356, doi:\href{https://doi.org/10.1016/j.physletb.2008.11.036}{\nolinkurl{10.1016/j.physletb.2008.11.036}}.

\bibitem{CapriGZ2010}
M. A. L. Capri, A. J. G\'omez, M. S. Guimaraes, V. E. R. Lemes, S. P. Sorella and D. G. Tedesco, \textit{Remark on the BRST symmetry in the Gribov-Zwanziger theory}, Phys. Rev. D \textbf{82}, 105019 (2010), arXiv:1009.4135, doi:\href{https://doi.org/10.1103/PhysRevD.82.105019}{\nolinkurl{10.1103/PhysRevD.82.105019}}.

\bibitem{CapriGZ2011}
M. A. L. Capri, A. J. G\'omez, M. S. Guimaraes, V. E. R. Lemes, S. P. Sorella and D. G. Tedesco, \textit{Renormalizability of the linearly broken formulation of the BRST symmetry in the presence of the Gribov horizon in Landau gauge Euclidean Yang-Mills theories}, Phys. Rev. D \textbf{83}, 105001 (2011), arXiv:1102.5695, doi:\href{https://doi.org/10.1103/PhysRevD.83.105001}{\nolinkurl{10.1103/PhysRevD.83.105001}}.

\bibitem{DudalSorella2012}
D. Dudal and S. P. Sorella, \textit{Gribov horizon and spontaneous BRST symmetry breaking}, Phys. Rev. D \textbf{86}, 045005 (2012), arXiv:1205.3934, doi:\href{https://doi.org/10.1103/PhysRevD.86.045005}{\nolinkurl{10.1103/PhysRevD.86.045005}}.

\bibitem{Capri2013}
M. A. L. Capri, D. Dudal, M. S. Guimaraes, I. F. Justo, L. F. Palhares and S. P. Sorella, \textit{Spontaneous breaking of the BRST symmetry in the presence of the Gribov horizon: Renormalizability}, Ann. Phys. \textbf{339}, 344-357 (2013), arXiv:1306.3122, doi:\href{https://doi.org/10.1016/j.aop.2013.09.006}{\nolinkurl{10.1016/j.aop.2013.09.006}}.

\bibitem{Capri2014}
M. A. L. Capri, M. S. Guimaraes, I. F. Justo, L. F. Palhares and S. P. Sorella, \textit{Properties of the Faddeev-Popov operator in the Landau gauge, matter confinement and soft BRST breaking}, Phys. Rev. D \textbf{90}, 085010 (2014), arXiv:1408.3597, doi:\href{https://doi.org/10.1103/PhysRevD.90.085010}{\nolinkurl{10.1103/PhysRevD.90.085010}}.

\bibitem{Capri2015}
M. A. L. Capri, D. Fiorentini, M. S. Guimaraes, B. W. Mintz, L. F. Palhares, S. P. Sorella, D. Dudal, I. F. Justo, A. D. Pereira and R. F. Sobreiro, \textit{Exact nilpotent nonperturbative BRST symmetry for the Gribov-Zwanziger action in the linear covariant gauge}, Phys. Rev. D \textbf{92}, 045039 (2015), arXiv:1506.06995, doi:\href{https://doi.org/10.1103/PhysRevD.92.045039}{\nolinkurl{10.1103/PhysRevD.92.045039}}.

\bibitem{Capri2016}
M. A. L. Capri, D. Dudal, D. Fiorentini, M. S. Guimaraes, I. F. Justo, A. D. Pereira, B. W. Mintz, L. F. Palhares, R. F. Sobreiro and S. P. Sorella, \textit{Local and BRST-invariant Yang-Mills theory within the Gribov horizon}, Phys. Rev. D \textbf{94}, 025035 (2016), arXiv:1605.02610, doi:\href{https://doi.org/10.1103/PhysRevD.94.025035}{\nolinkurl{10.1103/PhysRevD.94.025035}}.

\bibitem{Mintz2019}
B. W. Mintz, L. F. Palhares, G. Peruzzo and S. P. Sorella, \textit{Infrared massive gluon propagator from a BRST-invariant Gribov horizon in a family of covariant gauges}, Phys. Rev. D \textbf{99}, 034002 (2019), arXiv:1812.03166, doi:\href{https://doi.org/10.1103/PhysRevD.99.034002}{\nolinkurl{10.1103/PhysRevD.99.034002}}.

\bibitem{Sorella2009}
S. P. Sorella, \textit{Gribov horizon and BRST symmetry: a few remarks}, Phys. Rev. D \textbf{80}, 025013 (2009), arXiv:0905.1010, doi:\href{https://doi.org/10.1103/PhysRevD.80.025013}{\nolinkurl{10.1103/PhysRevD.80.025013}}.

\bibitem{Gracey2010}
J. A. Gracey, \textit{Alternative refined Gribov-Zwanziger Lagrangian}, Phys. Rev. D \textbf{82}, 085032 (2010), arXiv:1009.3889, doi:\href{https://doi.org/10.1103/PhysRevD.82.085032}{\nolinkurl{10.1103/PhysRevD.82.085032}}.

\bibitem{SchadenZwanziger1994}
M. Schaden and D. Zwanziger, \textit{Horizon condition holds pointwise on finite lattice with free boundary conditions}, in \textit{Proceedings of the Workshop on Quantum Infrared Physics}, Paris, 6-10 June 1994, arXiv:hep-th/9410019.

\bibitem{DudalVacuum2019}
D. Dudal, C. P. Felix, L. F. Palhares, F. Rondeau and D. Vercauteren, \textit{The BRST-invariant vacuum state of the Gribov-Zwanziger theory}, Eur. Phys. J. C \textbf{79}, 731 (2019), arXiv:1901.11264, doi:\href{https://doi.org/10.1140/epjc/s10052-019-7235-0}{\nolinkurl{10.1140/epjc/s10052-019-7235-0}}.

\bibitem{GuimaraesLorentzian2026}
M. S. Guimaraes, \textit{Lorentzian Gribov no-pole condition for Yang-Mills theory}, Phys. Rev. D \textbf{114}, 045024 (2026), arXiv:2606.08697, doi:\href{https://doi.org/10.1103/8jz8-v392}{\nolinkurl{10.1103/8jz8-v392}}.

\bibitem{SchadenZwanziger2015a}
M. Schaden and D. Zwanziger, \textit{Living with spontaneously broken BRST symmetry. I. Physical states and cohomology}, Phys. Rev. D \textbf{92}, 025001 (2015), arXiv:1412.4823, doi:\href{https://doi.org/10.1103/PhysRevD.92.025001}{\nolinkurl{10.1103/PhysRevD.92.025001}}.

\bibitem{SchadenZwanziger2015b}
M. Schaden and D. Zwanziger, \textit{Living with spontaneously broken BRST symmetry. II. Poincar\'e invariance}, Phys. Rev. D \textbf{92}, 025002 (2015), arXiv:1501.05974, doi:\href{https://doi.org/10.1103/PhysRevD.92.025002}{\nolinkurl{10.1103/PhysRevD.92.025002}}.

\bibitem{Gracey2006}
J. A. Gracey, \textit{Two loop correction to the Gribov mass gap equation in Landau gauge QCD}, Phys. Lett. B \textbf{632}, 282-286 (2006), Erratum Phys. Lett. B \textbf{686}, 319 (2010), arXiv:hep-ph/0510151, doi:\href{https://doi.org/10.1016/j.physletb.2005.10.046}{\nolinkurl{10.1016/j.physletb.2005.10.046}}; erratum doi:\href{https://doi.org/10.1016/j.physletb.2010.02.039}{\nolinkurl{10.1016/j.physletb.2010.02.039}}.

\bibitem{CucchieriFMR1998}
A. Cucchieri, \textit{Numerical study of the fundamental modular region in the minimal Landau gauge}, Nucl. Phys. B \textbf{521}, 365-379 (1998), arXiv:hep-lat/9711024, doi:\href{https://doi.org/10.1016/S0550-3213(98)00235-1}{\nolinkurl{10.1016/S0550-3213(98)00235-1}}.

\bibitem{CucchieriMendesGluon2008}
A. Cucchieri and T. Mendes, \textit{Constraints on the infrared behavior of the gluon propagator in Yang-Mills theories}, Phys. Rev. Lett. \textbf{100}, 241601 (2008), arXiv:0712.3517, doi:\href{https://doi.org/10.1103/PhysRevLett.100.241601}{\nolinkurl{10.1103/PhysRevLett.100.241601}}.

\bibitem{Bogolubsky2009}
I. L. Bogolubsky, E.-M. Ilgenfritz, M. M\"uller-Preussker and A. Sternbeck, \textit{Lattice gluodynamics computation of Landau gauge Green's functions in the deep infrared}, Phys. Lett. B \textbf{676}, 69-73 (2009), arXiv:0901.0736, doi:\href{https://doi.org/10.1016/j.physletb.2009.04.076}{\nolinkurl{10.1016/j.physletb.2009.04.076}}.

\bibitem{CucchieriMendesGhost2008}
A. Cucchieri and T. Mendes, \textit{Constraints on the infrared behavior of the ghost propagator in Yang-Mills theories}, Phys. Rev. D \textbf{78}, 094503 (2008), arXiv:0804.2371, doi:\href{https://doi.org/10.1103/PhysRevD.78.094503}{\nolinkurl{10.1103/PhysRevD.78.094503}}.

\bibitem{CucchieriMendes2013}
A. Cucchieri and T. Mendes, \textit{Ghost sector and geometry in minimal Landau gauge: further constraining the infinite-volume limit}, Phys. Rev. D \textbf{88}, 114501 (2013), arXiv:1308.1283, doi:\href{https://doi.org/10.1103/PhysRevD.88.114501}{\nolinkurl{10.1103/PhysRevD.88.114501}}.

\bibitem{DudalBounds2015}
D. Dudal, M. S. Guimaraes, I. F. Justo and S. P. Sorella, \textit{On bounds and boundary conditions in the continuum Landau gauge}, Eur. Phys. J. C \textbf{75}, 83 (2015), arXiv:1411.2500, doi:\href{https://doi.org/10.1140/epjc/s10052-015-3303-2}{\nolinkurl{10.1140/epjc/s10052-015-3303-2}}.

\bibitem{SternbeckMuller2013}
A. Sternbeck and M. M\"uller-Preussker, \textit{Lattice evidence for the family of decoupling solutions of Landau gauge Yang-Mills theory}, Phys. Lett. B \textbf{726}, 396-403 (2013), arXiv:1211.3057, doi:\href{https://doi.org/10.1016/j.physletb.2013.08.017}{\nolinkurl{10.1016/j.physletb.2013.08.017}}.

\bibitem{SternbeckFP2006}
A. Sternbeck, E.-M. Ilgenfritz and M. M\"uller-Preussker, \textit{Spectral properties of the Landau gauge Faddeev-Popov operator in lattice gluodynamics}, Phys. Rev. D \textbf{73}, 014502 (2006), arXiv:hep-lat/0510109, doi:\href{https://doi.org/10.1103/PhysRevD.73.014502}{\nolinkurl{10.1103/PhysRevD.73.014502}}.

\bibitem{Greensite2010}
J. Greensite, \textit{Faddeev-Popov spectra at the Gribov horizon}, Phys. Rev. D \textbf{81}, 114011 (2010), arXiv:1001.0784, doi:\href{https://doi.org/10.1103/PhysRevD.81.114011}{\nolinkurl{10.1103/PhysRevD.81.114011}}.

\bibitem{ArnoldFalkWinther2010}
D. N. Arnold, R. S. Falk and R. Winther, \textit{Finite element exterior calculus: from Hodge theory to numerical stability}, Bull. Am. Math. Soc. \textbf{47}, 281-354 (2010), arXiv:0906.4325, doi:\href{https://doi.org/10.1090/S0273-0979-10-01278-4}{\nolinkurl{10.1090/S0273-0979-10-01278-4}}.

\bibitem{DroniouYangMills2022}
J. Droniou, T. A. Oliynyk and J. J. Qian, \textit{A polyhedral discrete de Rham numerical scheme for the Yang-Mills equations}, J. Comput. Phys. \textbf{478}, 111955 (2023), arXiv:2208.12009, doi:\href{https://doi.org/10.1016/j.jcp.2023.111955}{\nolinkurl{10.1016/j.jcp.2023.111955}}.

\bibitem{TedescoDarkCrossings}
D. G. Tedesco, \textit{Blind Spots of the Zwanziger Horizon Function}, arXiv:2608.12413v2 (2026).

\end{thebibliography}
\end{document}